\documentclass[aps,twocolumn,prc,superscriptaddress,showpacs,nofootinbib,floatfix,amssymb,amsfonts,amsmath]{revtex4-1}
\usepackage{graphicx}
\usepackage{dcolumn}
\usepackage{bm}

\usepackage{amsmath}    
\usepackage{amsfonts}   
\usepackage{amssymb}
\usepackage{graphicx}   

\begin{document}

\title{Covariant energy density functionals: nuclear matter constraints
       and global ground state properties}

\author{A.\ V.\ Afanasjev}
\affiliation{Department of Physics and Astronomy, Mississippi
State University, MS 39762}

\author{S.\ E.\ Agbemava}
\affiliation{Department of Physics and Astronomy, Mississippi
State University, MS 39762}

\date{\today}

\begin{abstract}
   The correlations between global description of the ground state properties (binding 
energies, charge radii) and nuclear matter properties of the state-of-the-art 
covariant energy density functionals have been studied. It was concluded that 
the strict enforcement of the constraints on the nuclear matter properties (NMP) 
defined in Ref.\ \cite{RMF-nm} will not necessary lead to the functionals with good 
description of the binding energies and other ground and excited state properties.
In addition, it will not substantially reduce the uncertainties in the predictions of 
the binding energies in neutron-rich systems. It turns out that the functionals, which 
come close to satisfying these NMP constraints, have some problems in the description 
of existing data. On the other hand, these problems are either absent or much smaller in the 
functionals which are carefully fitted to finite nuclei but which violate some NMP 
constraints.  This is a consequence of the fact that the properties of finite nuclei 
are defined not only by nuclear matter properties but also by underlying shell effects.
The mismatch of phenomenological content, existing in all modern functionals, related 
to nuclear matter physics and the physics of finite nuclei could also be responsible.
\end{abstract}

\pacs{21.60.Jz, 21.10.Dr, 21.10.Ft, 21.10.Gv}

\maketitle

\section{INTRODUCTION}

   Bound states of the nucleons manifest themself in two species:
finite nuclei and neutron stars. The former system is bound by
strong forces, while the latter by gravitational ones. The description 
of both types of nuclear systems is intimately connected with
a concept of nuclear matter which is an idealized infinite system
of nucleons (neutrons and protons) interacting by strong 
forces. Infinite volume implies no surface effects and translational
invariance. This concept is well suited for the description of the
properties of interior of neutron stars. 

  However, it also has some important implications for finite nuclei 
(see Refs.\ \cite{NRSV.14,PCFG.14,P.14,CGS.14,HKM.14} in recent topical 
review on nuclear symmetry energy).  This is because the constraints 
on nuclear matter properties (NMP) enter into fitting protocols of
the energy density functionals (EDF) for non-relativistic and covariant
density functional theories \cite{BHP.03,VALR.05} (the abbreviation 
CDFT is used for latter one). In this way they affect the properties 
of finite nuclei (both static and dynamic aspects) 
\cite{BHP.03,VALR.05,NRSV.14,PCFG.14,P.14,CGS.14}.

  The analysis of the 263 covariant energy density functionals (further CEDFs) 
with respect of NMP constraints has recently been performed in Ref.\ \cite{RMF-nm}. 
Note that only small portion of these functionals (less than 10) have been used 
in a more or less systematic studies of the properties of finite nuclei. The 
performance of other functionals with respect of the description of finite nuclei 
(apart of few spherical nuclei used in the fitting protocols) is not known. The 
properties of symmetric nuclear matter, pure neutron matter, symmetry energy and 
its derivatives were constrained based on experimental/empirical data and model 
calculations in Ref.\ \cite{RMF-nm}. This resulted in two sets of constraints called  
SET2a and SET2b relevant for CDFT models; the part of these constraints is listed in 
Table \ref{Table-mass-NMP} below. Note that they are characterized by substantial 
uncertainties.

  It turns out that among these 263 CEDFs only 4  and 3 satisfy SET2a and SET2b 
NMP constraints, respectively. However, these functionals have never been used 
in the studies of finite nuclei. Thus, it is impossible to verify whether 
good NMP of these functionals will translate into good 
global description of binding energies, charge radii, deformations etc. 
Removing isospin incompressibility constraint increases the number of 
functionals which satisfy SET2a and SET2b constraints to 35 and 30, 
respectively \cite{RMF-nm}. Again the performance of absolute majority of 
these functionals in finite nuclei is not known. However, among those are 
the FSUGold and DD-ME$\delta$ CEDFs the global performance of which has been 
studied in the RMF+BCS and RHB models in Refs.\ \cite{RA.11,AARR.14}, 
respectively. Additional constraints on the functionals come from the properties 
of neutron stars \cite{DLM.16}. It turns out that FSUGold and DD-ME$\delta$ place 
maximum mass $M$ of neutron  star well below  and above the measured limit of 
$1.93 \leq M/M_{\odot} \leq 2.05$ 
\cite{NS_mass.1,NS_mass.2} where $M_{\odot}$ is the solar mass. The DD-ME$\delta$ 
functional comes to this limit only when hyperons are included; however, there 
are substantial uncertainties in the meson-hyperon couplings \cite{DLM.16}
as well as in the existence of hyperons in the interior of neutron stars 
\cite{CV.16}.

 Thus, the number of questions emerge.  First one is whether strict enforcement 
of these NMP constraints will inevitably lead to an improvement of the description 
of the ground state properties of finite nuclei in the CDFT and to a reduction of theoretical 
uncertainties in the description of the properties of neutron-rich nuclei.   
Another question is whether there is some physics missing in the current generation of CEDFs 
which could be responsible for some mismatch of the results for finite nuclei 
and neutron stars. It is also important to understand how the details of the
fitting protocols affect these conclusions. The study of these questions
represent the first goal of the present paper. The second goal is to understand 
whether future experimental data on the ground state properties of neutron-rich 
nuclei will allow to minimize theoretical 
uncertainties for physical observables of  neutron-rich nuclei and to which extent.
Such an approach assumes more reliance on the data on finite nuclei and less
dependence on the NMP constraints in the definition of isovector properties
of CEDFs.

 To address these questions we perform the global analysis of the  ground 
state observables such as binding energies and charge radii obtained with the 
state-of-the-art CEDFs which differ substantially in the NMPs. CEDF DD-ME$\delta$, 
which is coming very close to satisfying all required NMP constraints, is among 
them. Binding energies of finite nuclei play an important role in the nuclear
structure and nuclear astrophysics. Their evolution with charge and isospin 
defines the limits of nuclear landscape (Refs.\ \cite{Eet.12,AARR.13,AARR.14}).
Accurate modeling of nuclear astrophysics processes and the reduction of 
relevant theoretical uncertainties requires the precise knowledge of binding
energies of neutron-rich nuclei which are currently non-accessible by experimental
facilities \cite{MSMA.16}.

  The paper is organized as follows. Sec.\ \ref{theory} presents brief outline of 
theoretical framework. Theoretical uncertainties in the predictions of binding energies
and the role of future experimental facilities in their reduction are 
discussed in Sec.\ \ref{th-uncert}. Sec.\ \ref{sec-impact} considers the 
impact of nuclear matter properties of the functionals on the predictions
of binding energies in known and neutron-rich nuclei.
The accuracy of the 
description of the ground state properties of finite nuclei and its dependence 
on fitting protocol are discussed in Sec.\ \ref{sec-finite}.  Sec.\ 
\ref{sec-general} is devoted to general observations following from this 
study. Finally, Sec.\ \ref{concl} summarizes the  results of our work.

\section{Brief outline of the details of theoretical framework}
\label{theory}

    The results have been obtained in the relativistic Hartree-Bogoliubov 
(RHB) framework the details of which are discussed in Secs.\ III and IV of 
Ref.\ \cite{AARR.14}. These deformed RHB calculations are restricted to axial
reflection symmetric shapes.

  We focus on four CEDFs (NL3* \cite{NL3*}, DD-ME2 \cite{DD-ME2}, DD-PC1 \cite{DD-PC1} 
and DD-ME$\delta$ \cite{DD-MEdelta}) which were used in the global studies 
of Refs.\ \cite{AARR.13,AARR.14,AARR.15} and for which numerical results 
are available. These functionals are compared in Sec.\ 2 of  Ref.\ \cite{AARR.14}.
To deal with complete set of major classes of the state-of-the-art CEDFs, we 
also provide new results obtained with CEDF PC-PK1 \cite{PC-PK1} for the Yb 
isotope chain.  This functional has been used with success 
for the studies of the masses of known nuclei by Peking group in Refs.\ 
\cite{ZNLYM.14,LLLYM.15}. 

   These functionals reproduce the binding energies of known nuclei at the
mean field level with the rms-deviations of around 2.5 MeV (see Table 
\ref{rms-fit-global2} below). However, they differ substantially in the 
underlying physics (see discussion in Sec.\ 2 of Ref.\ \cite{AARR.14}
and Ref.\ \cite{PC-PK1}) and fitting protocols (see Table \ref{table-fit} 
and Fig.\ \ref{E_fitted} below).

\begin{table}[h]
  \begin{center}
  \caption{Input data for fitting protocols of different CEDFs. 
           Columns (2-4) show the number of experimental data 
           points on binding energies $E$, charge radii $r_{ch}$ 
           and neutron skin thicknesses  $r_{skin}$ used in the 
           fitting protocols. Column 5 indicates which type of 
           nuclei (spherical (S) or deformed (D)) were used. 
           Column 6 shows whether microscopic equation of state
           (EOS) has been used in the fit of the functional or not;
           here ``Y'' stands for ``yes'' and ``N'' for ``no''.
\label{table-fit}
}
  \begin{tabular}{|c|c|c|c|c|c|} \hline
     CEDF       &  $E$   &  $r_{ch}$  &  $r_{skin}$  &  Type of nuclei & EOS \\ \hline
       1        &   2   &  3        &   4         &   5             & 6   \\ \hline
     NL3*       &  12   &  9        &   4         &   S             & N   \\
    DD-ME2      &  12   &  9        &   3         &   S             & N   \\
  DD-ME$\delta$ &  161  &  86       &   0         &   S             & Y   \\
    DD-PC1      &  64   &  0        &   0         &   D             & Y   \\
    PC-PK1      &  60   &  17       &   0         &   S             & N   \\
  \hline
  \end{tabular}
  \end{center}
\end{table}

 Table \ref{table-fit} shows that only two of these functionals, namely,
DD-ME$\delta$ and DD-PC1, are fitted to the equation of state (EOS) 
of  neutron matter obtained in microscopic calculations with realistic 
forces. Although these EOS are similar at saturation densities, they
differ substantially in their stifness at the densities typical to the
centre of neutron stars \cite{APR.98,LS.08,PCFG.14}. Note that no 
reliable data, either observational or experimental, exist for such 
densities. As a result, there is no way to discriminate these predictions
for the EOS.

\begin{figure*}[ht]
\includegraphics[angle=0,width=11.8cm]{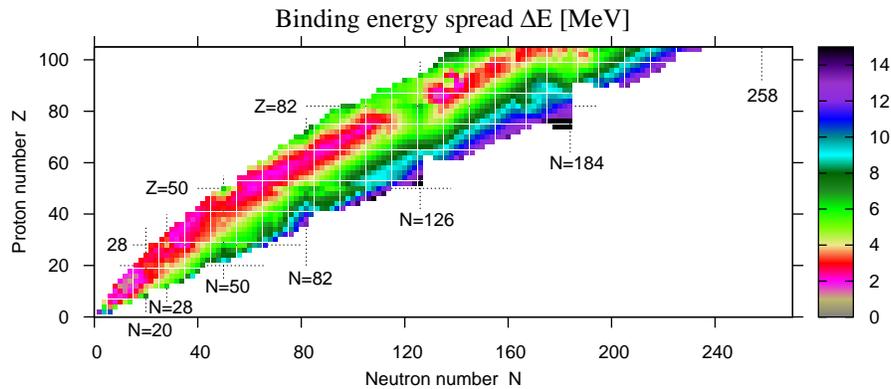}
\caption{(Color online) The binding energy spreads $\Delta E(Z,N)$ as 
a function of proton and neutron number. From Ref.\ \cite{AARR.14}.}
\label{DE_spread_total}
\end{figure*}

\begin{table*}[ht]
  \begin{center}
  \caption{The rms deviations $\Delta{E_{rms}}$ and $\Delta{(r_{ch})_{rms}}$ 
           between calculated and experimental binding energies $E$ and charge 
           radii $r_{ch}$. The columns (2) and (4) show the $\Delta{E_{rms}}$ 
           and $\Delta{(r_{ch})_{rms}}$ values obtained in the RHB calculations 
           of Ref.\ \cite{AARR.14} for the nuclei used in the fitting protocols 
           of first four functionals. Note that our results for the 
           DD-ME$\delta$ differ from the ones obtained in Ref.\ \cite{DD-MEdelta}
           which are shown in brackets. This is because 24 out of 161 nuclei
           used in the fit are deformed in our calculations; note that all these
           161 nuclei were assumed to be spherical in Ref.\ \cite{DD-MEdelta}.
           The results for CEDF PC-PK1 are from Ref.\ 
           \cite{PC-PK1}. The columns (3) and (5) show the $\Delta E_{rms}$
           and $\Delta{(r_{ch})_{rms}}$ values obtained in global calculations. 
           For first four functionals, they are defined in Ref.\ \cite{AARR.14} 
           with respect of 640 measured masses presented in the AME2012 
           compilation \cite{AME2012}. For PC-PK1 they are defined with respect 
           of 575 masses in Ref.\ \cite{ZNLYM.14}.
  \label{rms-fit-global2}
}
  \begin{tabular}{|c|c|c|c|c|} \hline
   CEDF & $\Delta E_{rms}^{fit}$ [MeV] & $\Delta E_{rms}^{global}$ [MeV]
   & $\Delta (r_{ch})_{rms}^{fit} [$fm$] $ & $\Delta (r_{ch})_{rms}^{global}$ [$fm$] \\ 
   \hline
   1  &   2  &  3 & 4 & 5 \\ \hline
    NL3*        & 1.68        & 2.96 & 0.017        & 0.0283 \\
    DD-ME2      & 1.48        & 2.39 & 0.015        & 0.0230 \\
  DD-ME$\delta$ & 2.33 [2.4]  & 2.29 & 0.028 [0.02] & 0.0329 \\
    DD-PC1      & 0.69        & 2.01 & 0.039\footnote{Note that no information
           on charge radii has been used in the fit of the DD-PC1 CEDF 
           \cite{DD-PC1}.} & 0.0253 \\
    PC-PK1      & 1.33 & 2.58 & 0.019 &        \\
  \hline
  \end{tabular}
  \end{center}
\end{table*}

\begin{figure*}[ht]
\includegraphics[angle=-90,width=8.8cm]{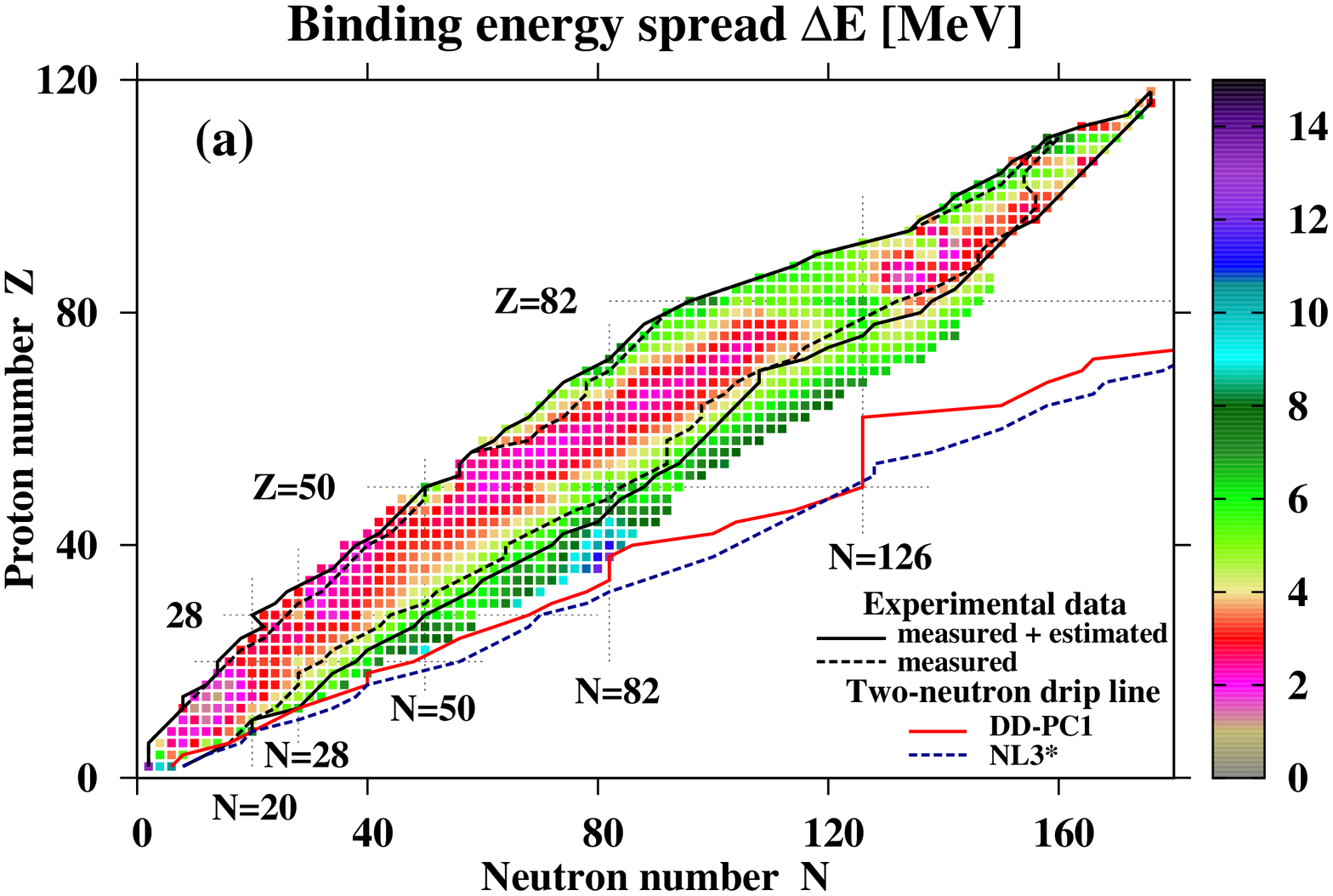}
\includegraphics[angle=-90,width=8.8cm]{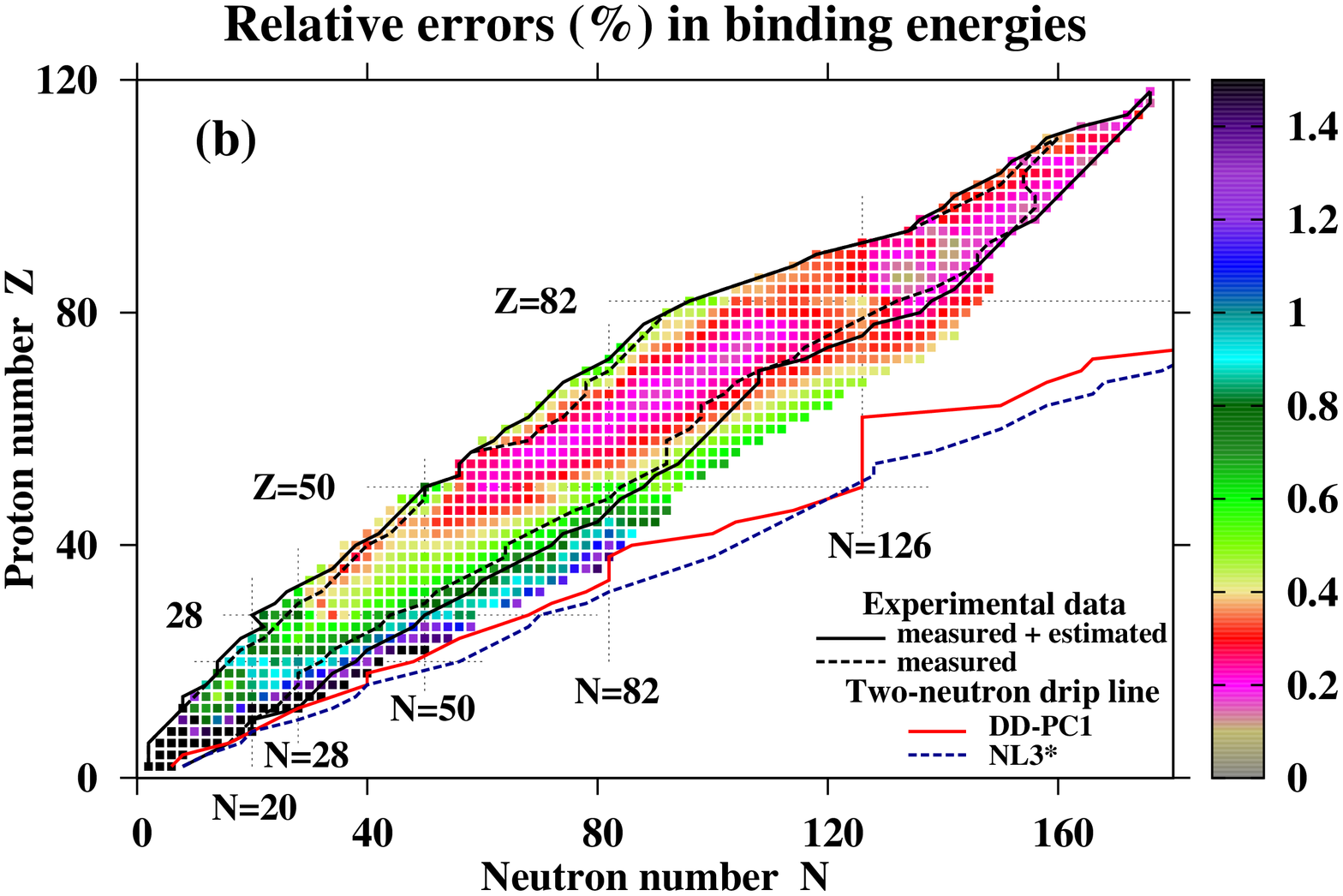}
\caption{ (Color online) Panel (a) is based on the results presented 
in Fig.\ \ref{DE_spread_total}. However, the squares are shown only for the 
nuclei which are currently known and which will be measured with FRIB. 
The regions of the nuclei with measured and measured+estimated masses 
are enclosed by dashed and solid black lines, respectively. The squares
beyond these regions indicate the nuclei which may be measured with FRIB.
For simplicity the line formed by most neutron-rich nucleus in each isotope 
chain accessible with FRIB will be called as ``FRIB limit''.
The FRIB limit of the nuclear chart is defined by fission yield greater 
than $10^{-6}$ that may be achieved with dedicated existence measurements 
\cite{S-priv.14}. The same colormap as in Fig.\ \ref{DE_spread_total} is used 
here, but the ranges of particle numbers for vertical and horizontal axis are 
different from the ones in Fig.\ \ref{DE_spread_total}. The two-neutron drip 
lines are shown for the CEDFs NL3* 
and DD-PC1 by blue dashed and solid red lines, respectively. Panel (b) is based on the
same results as panel (a), but the spreads in relative errors in the description of 
the masses are shown instead of binding energy spreads $\Delta E$.
} 
\label{FRIB-data-constraint}
\end{figure*}

\section{Theoretical uncertainties in the predictions of binding
         energies and the role of future experimental facilities
         in their reduction}
\label{th-uncert}

  The map of theoretical spreads $\Delta E(Z,N)$ in the predictions 
of the binding energies is shown in Fig.\ \ref{DE_spread_total}. These
spreads are defined as
\begin{eqnarray}
\Delta E(Z,N) = |E_{max}(Z,N) - E_{min}(Z,N)|, 
\end{eqnarray}
where $E_{max}(Z,N)$ and $E_{min}(Z,N)$ are the largest and the smallest 
binding energies for each $(N,Z)$ nucleus obtained with four state-of-the-art 
functionals, namely, NL3*, DD-ME2, DD-ME$\delta$ and DD-PC1. Here
the results of the calculations of Ref.\ \cite{AARR.14}, covering
nuclear landscape between the two-proton and two-neutron drip lines, 
are used.  The
accuracy of the description of experimental masses by these functionals
is given in Table \ref{rms-fit-global2}.
Fig.\ \ref{FRIB-data-constraint}a shows 
that the spreads in the predictions 
of binding energies stay within 5-6 MeV for the known nuclei (the regions
with measured and measured+estimated masses\footnote{The masses given in the 
AME2012 mass evaluation \cite{AME2012} can be separated into two groups: 
One represents nuclei with masses defined only from experimental data, 
the other contains nuclei with masses depending in addition on either
interpolation or extrapolation procedures. For simplicity, we call the masses 
of the nuclei in the first and second groups as measured and estimated. There 
are 640 measured and 195 estimated masses of even-even nuclei in the AME2012 
mass evaluation.} in Fig.\ \ref{FRIB-data-constraint}a). 
These spreads are even smaller (typically around 3 MeV) for the nuclei in the 
valley of beta-stability. However, theoretical uncertainties for the masses 
increase drastically when approaching the neutron-drip line and in some nuclei
they reach 15 MeV. This is a consequence of poorly defined isovector properties 
of many CEDFs.

 Fig.\ \ref{FRIB-data-constraint}b shows the spreads of the relative 
errors in the predictions of binding energies which are defined as
\begin{eqnarray}
\Delta E^{rel}(Z,N) = \frac{|E_{max}(Z,N) - E_{min}(Z,N)|}{\frac{1}{4} 
\sum\limits_{i=1}^4 E_i(Z,N)} 
\end{eqnarray}
where 
$E_i(Z,N)$ is the binding energy obtained with the $i$-th functional. The
quantity in the denominator is the average binding energy of the $(Z,N)$ 
nucleus obtained with four CEDFs. These 
spreads in relative errors are largest in light nuclei due to the effects 
which are not taken into account at the DFT level (see Ref.\ \cite{AARR.14}). 
For known nuclei they gradually decrease with the increase of mass so that 
for the $A\geq 80$ nuclei the spreads in relative errors stay safely below 
$0.5\%$.

 It is important to understand how future mass measurements with rare 
isotope facilities (such as FRIB, GANIL, RIKEN and FAIR) could help to 
improve isovector properties of the functionals. Fig.\ \ref{FRIB-data-constraint} 
clearly shows that the increase of the neutron number beyond the region of known nuclei
leads to an increase of the $\Delta E$ and $\Delta E^{rel}$ spreads. However, 
apart of the $Z\sim 40, N\sim 82$ region these increases are quite modest 
in terms of binding energy spreads $\Delta E$ on going from currently known 
limit of neutron-rich nuclei (for which $\Delta E \sim 6$ MeV) up to the FRIB limit 
(for which $\Delta E \sim 8$ MeV). Note that for $Z\geq 70$ nuclei a similar 
transition almost does not increase the  $\Delta E$ and $\Delta E^{rel}$ 
spreads. The largest increase in the $\Delta E$ spreads is observed in the 
$Z\sim 40$ nuclei for which the transition from the limit of currently known
nuclei to the FRIB limit changes $\Delta E$ from $\sim 6$ MeV to $\sim 12$ 
MeV (Fig.\ \ref{FRIB-data-constraint}a).

\begin{figure}[ht]
\includegraphics[angle=0,width=8.6cm]{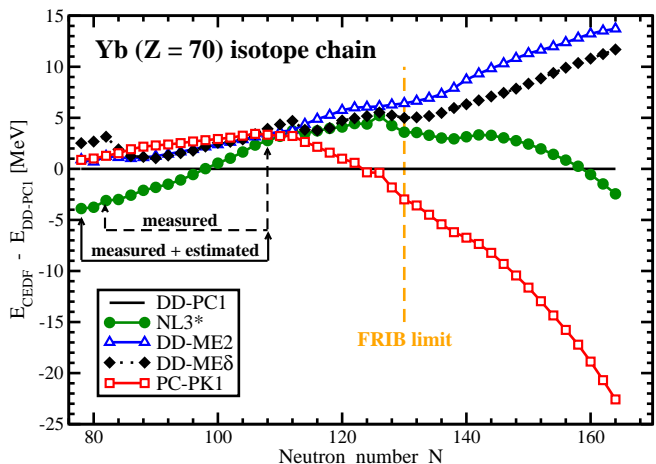}
\caption{(Color online) The differences in the calculated binding
energies  of the Yb nuclei
 obtained in the RHB calculations with two functionals.
The results obtained with DD-PC1 are used as a reference. The
regions of measured and measured+estimated masses from Ref.\ 
\cite{AME2012} are indicated by arrows. The FRIB limit is 
shown by dashed orange line.}
\label{Yb-with-PCPK1}
\end{figure}

 These results suggest that new mass measurements with future rare isotope 
beam facilities, which (dependent on model) may reach two-neutron drip line 
or its vicinity for the $Z\leq 26$ nuclei (Fig.\ \ref{FRIB-data-constraint}), 
could help to improve isovector properties of the CEDFs in the $Z\leq 50$ nuclei.  
However, this improvement is expected to be modest\footnote{This is in line 
with recent results of 
Ref.\ \cite{MSHSWN.15} which indicates that new mass measurements do not 
impose a strong enough constraint to generate significant changes in the 
energy density functionals.}. This is in part due to the fact that beyond 
mean field effects are quite 
important in light nuclei \cite{AARR.14} which complicates the use of future 
data on masses for the refit of the functionals at the DFT level.  The fact 
that the spread of relative errors in the predictions of the masses is the 
largest in light nuclei (see Fig.\ \ref{FRIB-data-constraint}b) also underlines 
the fact that light nuclei are less ``mean-field like'' as compared with
heavy ones. Even smaller improvement in the definition of isovector properties 
of the functionals is expected for the $50 \leq Z < 82$ nuclei since the increase 
in the $\Delta E$ spreads on going from the neutron-rich limit of the region
of known nuclei to the FRIB limit is rather small being typically around 2 MeV 
or less. It is also clear that future rare isotope beam facilities will contribute 
very little to a better understanding of the isovector properties of the $Z \geq 82$ 
nuclei (Fig.\ \ref{FRIB-data-constraint}).

 The results presented in Figs.\ \ref{DE_spread_total} and \ref{FRIB-data-constraint} 
are limited to the four CEDFs which were used in the global studies of Ref.\ \cite{AARR.14}. 
In general, this group of CEDFs has to be supplemented by the PC-PK1 functional since then 
the set of the state-of-the-art functional representing major classes of CEDFs will be 
complete. This has not been done in Ref.\ \cite{AARR.14} since the description of the 
ground state properties by PC-PK1 has been studied by the Peking group in the RMF+BCS 
framework in Ref.\ \cite{ZNLYM.14}. However, the properties of superheavy and octupole 
deformed nuclei have been studied globally with all five functionals in Refs.\ 
\cite{AANR.15,AAR.16}. 

\begin{figure*}[ht]
\includegraphics[angle=0,width=11.8cm]{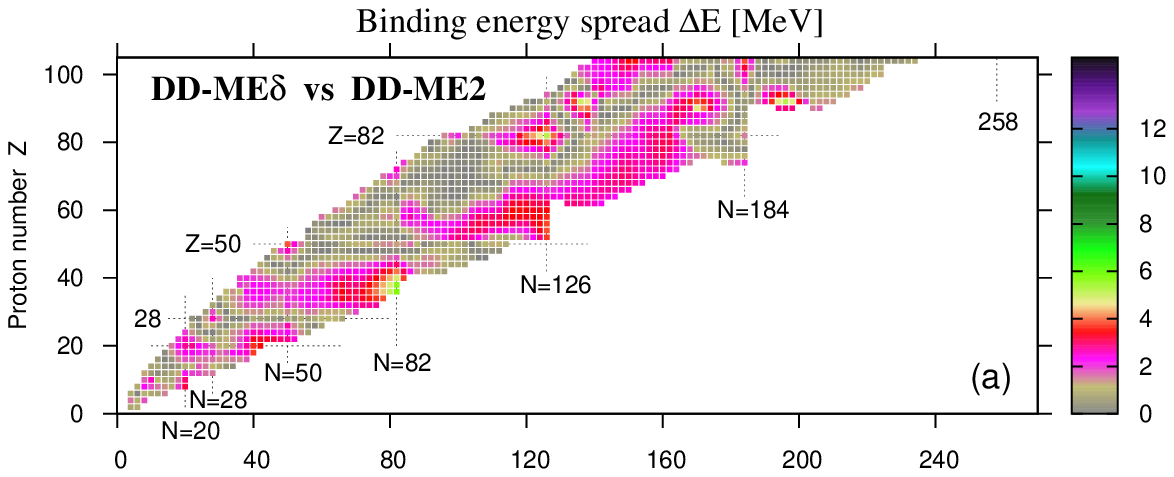}
\includegraphics[angle=0,width=11.8cm]{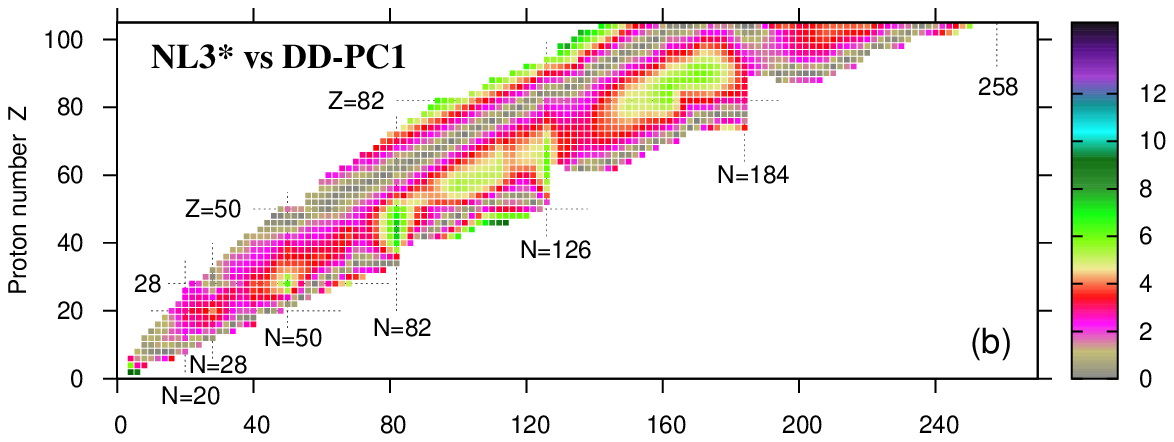}
\includegraphics[angle=0,width=11.8cm]{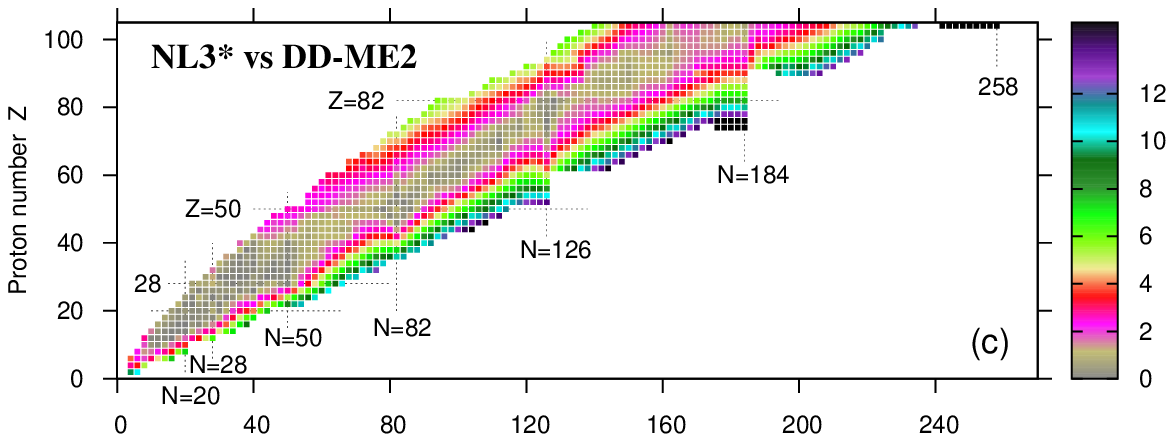}
\includegraphics[angle=0,width=11.8cm]{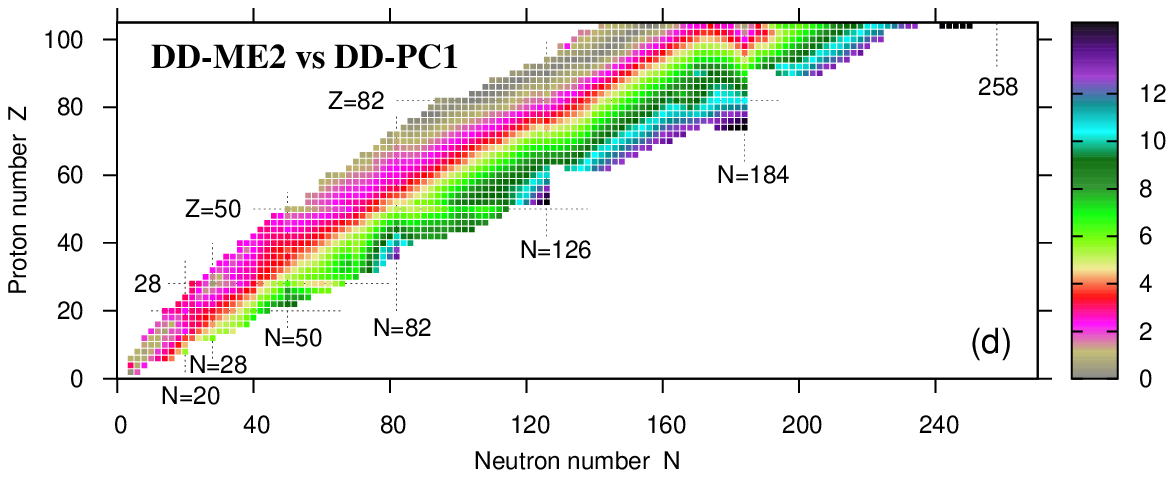}
\caption{(Color online) Binding energy spreads $\Delta E(Z,N)$ for the 
pairs of indicated functionals. All even-even nuclei between the two-proton
and two-neutron drip lines are included in the comparison.}
\label{E_spreads}
\end{figure*}

  We have not performed additional global studies with the PC-PK1 functional since 
the main conclusions can be derived based on the RHB results for the  CEDFs NL3*, DD-ME2, 
DD-PC1 and DD-ME$\delta$ obtained earlier in Ref.\ \cite{AARR.14}. However, we 
will illustrate the performance of this functional on the example of the Yb nuclei. The 
results for calculated binding energies are presented in Fig.\ \ref{Yb-with-PCPK1}. 
Here the results obtained with DD-PC1 are used as a reference since this functional
provides the best description of the binding energies (see Table 
\ref{rms-fit-global2} and Fig.\ \ref{E_th_vs_exp}d below). In the region of known 
nuclei, the predictions of the PC-PK1 functional are very close to the ones obtained with 
the DD-ME2 and DD-ME$\delta$ functionals. With increasing neutron number $N$ the PC-PK1 results come 
closer to those obtained with DD-PC1 and above $N=122$ the nuclei are more bound in PC-PK1 than in 
DD-PC1. However, for the Yb nuclei within the FRIB limit (apart of the $N=128, 130$ Yb isotopes), 
the spreads of binding energies $\Delta E$ presented in Figs.\ 
\ref{DE_spread_total} and  \ref{FRIB-data-constraint} 
would be only marginally affected by the addition of the results obtained with the PC-PK1 
functional. On the other hand, the effect of addition of PC-PK1 on $\Delta E$ is very 
substantial for the nuclei beyond the FRIB limit where $\Delta E$ increases from $\sim 16$ 
MeV to $\sim 37$ MeV for $N=162$ (see Fig.\ \ref{Yb-with-PCPK1}). These differences
are caused by different isovector properties of CEDFs under study since calculated
deformations are similar for all functionals. These results for $\Delta E$ are not 
surprising considering that even the functionals accurately fitted to the masses show 
substantial differences in the binding energies of neutron-rich nuclei. For example,
the Skyrme functionals HFB-22 and HFB-24, which describe known masses with an accuracy
of approximately 0.6 MeV \cite{PCFG.14}, are characterized by the $\Delta E$ values 
reaching 10 MeV in neutron-rich nuclei (see Fig.\ 3 in \cite{PCFG.14}). 

 Note that there are some important similarities between the NL3* and PC-PK1 
functionals. First, the evolution of the pairing energies and pairing gaps with neutron 
number in PC-PK1 is very similar to the one in NL3* (see Fig. 3 in Ref.\ \cite{AARR.15} 
for the NL3* results). Second, both functionals predict the position of the two-neutron 
drip line at higher neutron number (at $N=178$ for NL3* and at $N=184$ for PC-PK1) as
compared with the DD-* functionals with explicit density dependence which
predict it at $N\sim 162$ (see Table IV in Ref.\ \cite{AARR.14}).

\section{The impact of nuclear matter properties of the functionals on
the predictions of binding energies of known and neutron-rich nuclei}
\label{sec-impact}
  
 Although new experimental data on masses of neutron-rich nuclei 
generated by future rare isotope facilities will allow to improve 
the isovector properties of the energy density functionals, it is 
not likely that such an improvement will either eliminate or 
substantially reduce all possible uncertainties.
Moreover, it is not clear whether the bias towards light and medium 
mass nuclei generated by future experimental data could be avoided since very 
little extension of the nuclear chart will be generated for the $Z\geq 82$ 
nuclei by these experiments (Fig.\ \ref{FRIB-data-constraint}). This is 
precisely the region where most of unknown $Z\leq 120$ nuclei are located 
and where the distance (in terms of neutrons) between the region of known 
nuclei and the two-neutron drip line is the largest (see Fig. 1 in Ref.\ 
\cite{AARR.13}).

  The fitting protocols of EDFs always contain data on finite nuclei 
(typically binding energies, charge radii and occasionally neutron skin 
thicknesses) and pseudodata on NMP (see Table \ref{table-fit} and Sect.\ II 
in Ref.\ \cite{AARR.14} for more details).  Binding energies and radii show different 
sensitivity to various terms of the CEDFs and, in addition, there are some important 
correlations between the NMP and surface properties of the functionals. For example, 
the calculated 
binding energies are not very sensitive to the nuclear matter saturation density 
but are strongly influenced by the choice of the parameters which define the 
surface energy coefficient $a_s$ in the empirical mass formula \cite{DD-PC1}. 
Strong converse relation exists between the nuclear charge radii and the saturation 
density of symmetric nuclear matter $\rho_0$ \cite{RN.16}. In addition, there is 
a strong correlation between the slope of symmetry energy $L_0$ and neutron skins 
\cite{FP.11,RNC.15,RN.16} (see Refs.\ \cite{RN.16,RN.10,FP.11,RNC.15,PACNPRRV.12} for 
the discussion  of other correlations).

Considering that existing data 
on binding energies does not allow to fully establish isovector properties 
of EDFs and make reliable predictions for masses of neutron-rich 
nuclei, it is important to have a closer look on NMP in order to see 
whether strict enforcement of NMP constraints could reduce theoretical 
uncertainties in isovector properties of EDFs and mass predictions for 
neutron-rich nuclei.

\begin{table*}[ht]
\caption{Properties of symmetric nuclear matter at saturation: the density $\rho_0$, the energy per 
particle $E/A$, the incompressibility $K_0$, the symmetry energy $J$ and its slope $L_0$, and 
the Lorentz effective mass $m$*/$m$ \cite{JM.89} of a nucleon at the Fermi  surface. Top five 
lines show the values for indicated covariant energy density functionals, while bottom two lines
show two sets (SET2a and SET2b) of the constraints on the experimental/empirical ranges for 
the quantities of interest defined in Ref.\ \cite{RMF-nm}. The CEDF values which are located beyond 
the limits of the SET2b constraint set are shown in bold. }
\label{tab-nuclear-matter}
\begin{center}
\begin{tabular}{|c|c|c|c|c|c|c|}\hline
CEDF              & $\rho_0$ [fm$^{-3}$] & $E/A$ [MeV]       & $K_0$ [MeV]      & $J$ [MeV]     & $L_0$ [MeV]  &  $m$*/$m$   \\ \hline
      1                &       2        &       3           &      4           &      5        &    6         &   7      \\ \hline
NL3* \cite{NL3*}       &   0.150        &  -16.31           &    258           & {\bf 38.68}   & {\bf 122.6}  &  0.67    \\
DD-ME2 \cite{DD-ME2}   &  0.152         & -16.14            &    251           & 32.40         &  49.4        & 0.66     \\
DD-ME$\delta$ \cite{DD-MEdelta} & 0.152 & -16.12            &    219           & 32.35         &  52.9        & 0.61     \\
DD-PC1 \cite{DD-PC1,PC-PK1}&   0.152    & -16.06            &    230           & 33.00         &  68.4        & 0.66     \\
PC-PK1 \cite{PC-PK1}   &   0.154        & -16.12            &    238           & {\bf 35.6}    &  {\bf 113}   & 0.65     \\ 
 SET2a                 &  $\sim 0.15$   &  $\sim -16$       &  190-270         &  25-35        &   25-115     &          \\ 
 SET2b                 &  $\sim 0.15$   &  $\sim -16$       &  190-270         &  30-35        &   30-80      &          \\ \hline
\end{tabular}
\end{center}
\label{Table-mass-NMP}
\end{table*}

 One way to do that is to see whether there is one-to-one correspondence 
between the differences in NMP of two functionals and the differences in 
their description of binding energies. Fig.\ \ref{E_spreads} and Table 
\ref{Table-mass-NMP} are created for such an analysis. The differences 
of the binding energies of several pairs of CEDFs are compared in Fig.\ 
\ref{E_spreads}; they are based on the results of the RHB calculations
obtained in Ref.\ \cite{AARR.14}. Table \ref{Table-mass-NMP} summarizes 
the NMPs of employed functionals and the experimental/empirical ranges 
for the quantities of interest obtained in Ref.\ \cite{RMF-nm}. The 
binding energy per particle $E/A \sim -16$ MeV and the saturation density 
$\rho_0 \sim 0.15$ fm$^{-3}$ represent well established properties of
infinite nuclear matter. On the other hand, the incompressibility $K_0$ 
of symmetric nuclear matter, its symmetry energy $J$ and the slope 
$L_0$ of symmetry energy at saturation density are characterized by 
substantial uncertainties (see Ref.\ \cite{RMF-nm} for details).
Effective mass of the nucleon at the Fermi surface $m$*/$m$ 
is also poorly defined in experiment.

The smallest difference in the predictions of binding energies exists 
for the DD-ME2/DD-ME$\delta$ pair of the functionals (Fig.\ \ref{E_spreads}a); 
for almost half of the $Z\leq 104$ nuclear landscape their predictions 
differ by less than 1.5 MeV and only in a few points of nuclear landscape 
the differences in binding energies of two functionals are close to 5 MeV.
The NMPs of these two functionals are similar with some minor differences 
existing only for the incompressibility $K_{0}$ and Lorentz effective mass 
$m$*/$m$ (Table \ref{Table-mass-NMP}). However, the similarity of NMP does 
not necessarily lead to similar predictions of binding energies. This is 
illustrated in Fig.\ \ref{E_spreads}d on the example of the pair of 
the DD-ME2 and DD-PC1 functionals for which substantial differences in 
the predictions exist for quite similar NMP (Table \ref{Table-mass-NMP}).

  Even more striking example is seen in Fig.\ \ref{E_spreads}b where 
the NL3*/DD-PC1 pair of the functionals, which are characterized by a 
substantial differences in the energy per particle ($E/A$), symmetry energy 
$J$ and its slope $L_0$ (Table \ref{Table-mass-NMP}), have significantly 
smaller differences in predicted binding energies as compared with above 
mentioned  DD-ME2/DD-PC1 pair of the functionals. This is a consequence
of a peculiar feature of the relative behavior of the binding energies of
the NL3* and DD-PC1 functionals with increasing isospin which is clearly 
visible in Fig.\ \ref{Yb-with-PCPK1}.  Note that the $J$ and $L_0$
values of the NL3* functional are located outside the experimental/empirical 
ranges for these values defined in Ref.\ \cite{RMF-nm} (see Table 
\ref{Table-mass-NMP}). 

\begin{figure*}[ht]
\includegraphics[angle=-90,width=8.8cm]{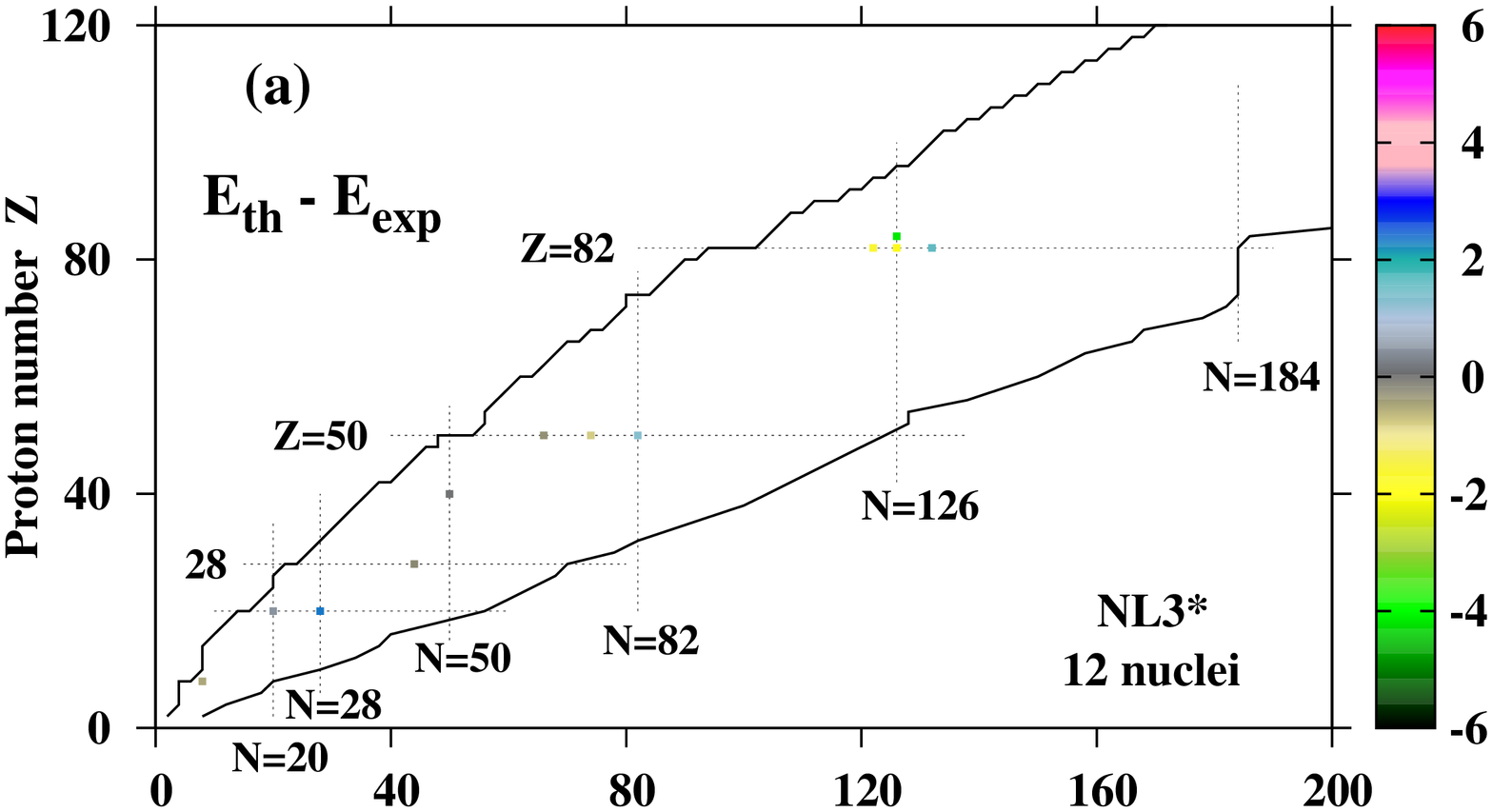}
\includegraphics[angle=-90,width=8.8cm]{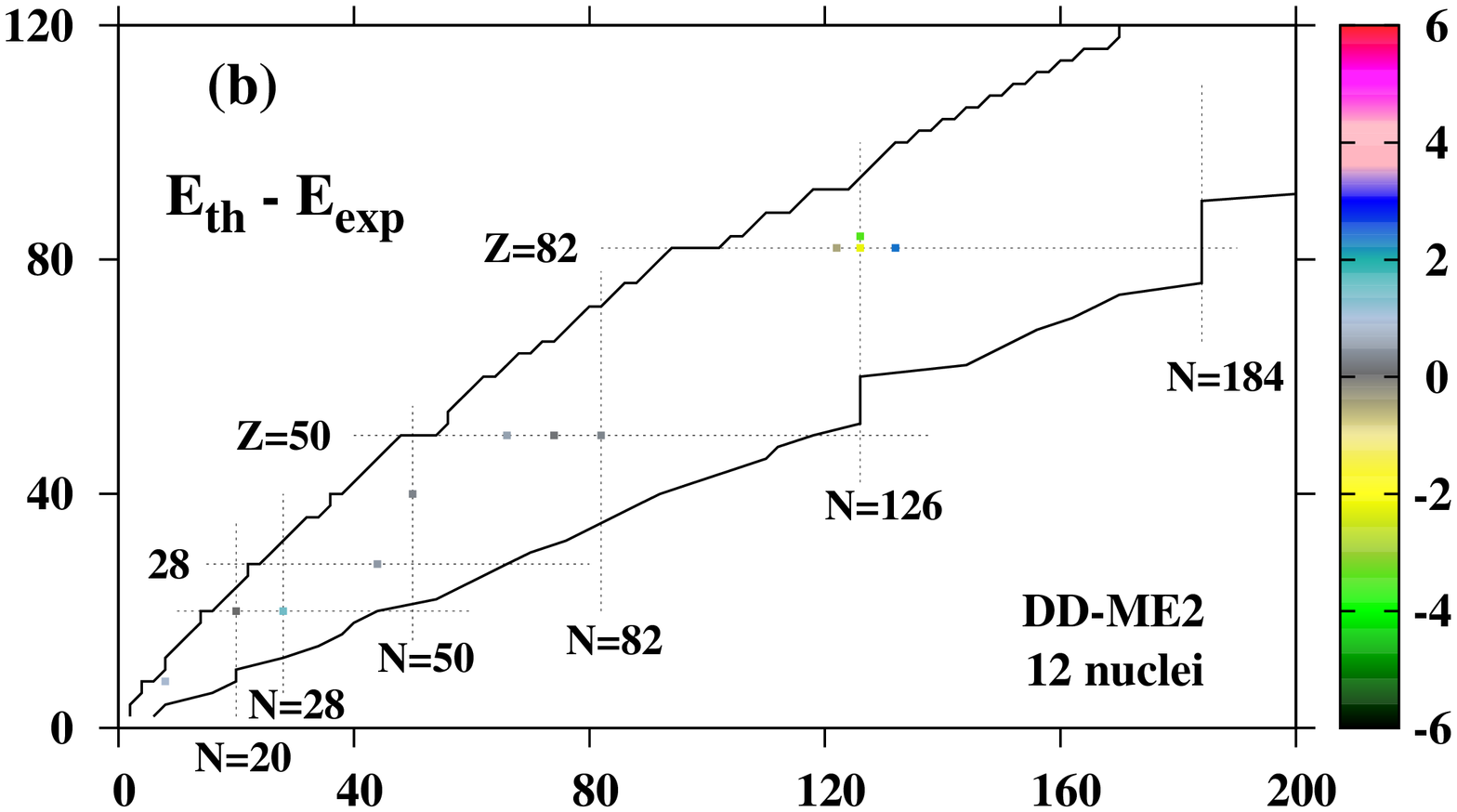}
\includegraphics[angle=-90,width=8.8cm]{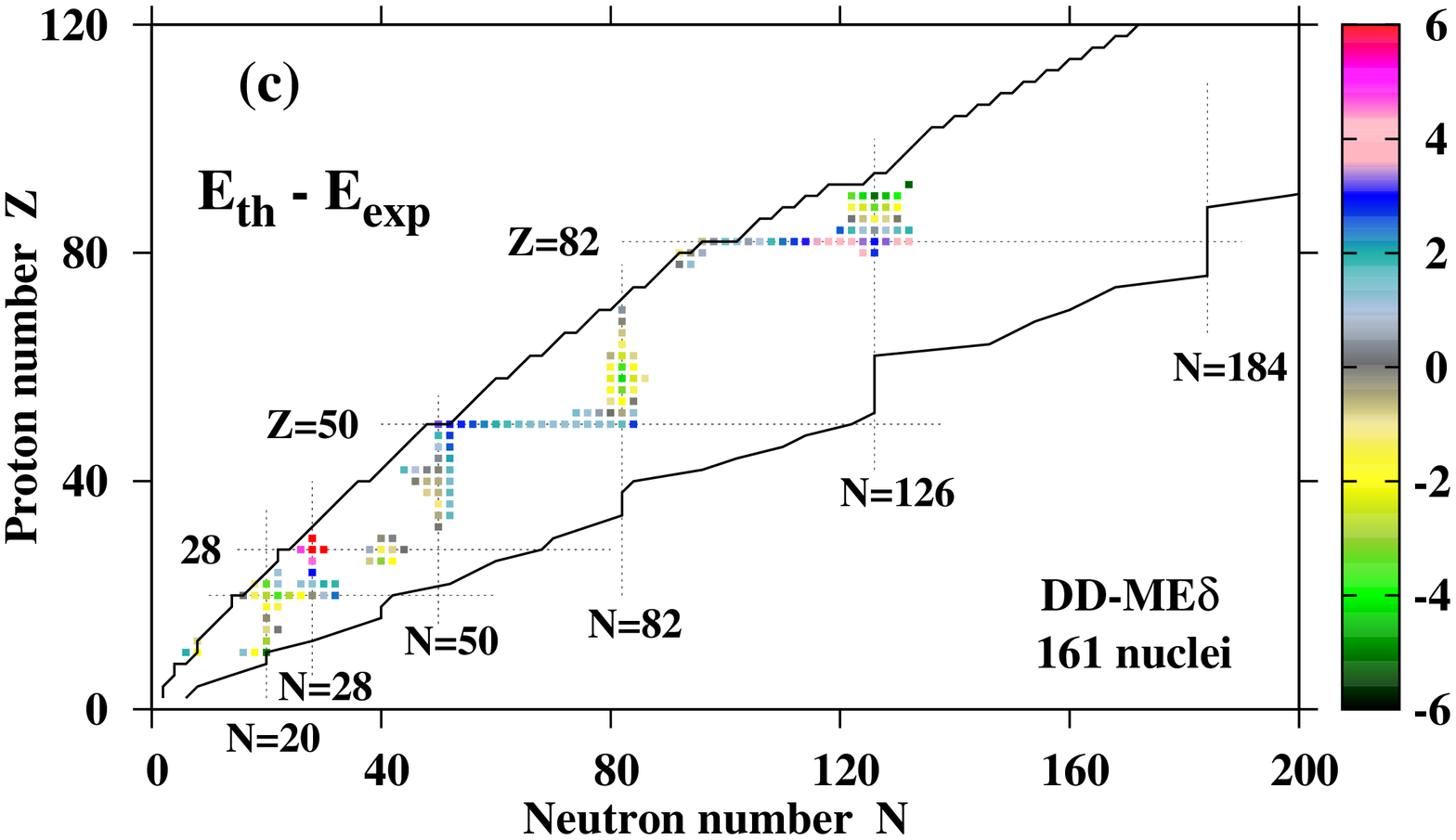}
\includegraphics[angle=-90,width=8.8cm]{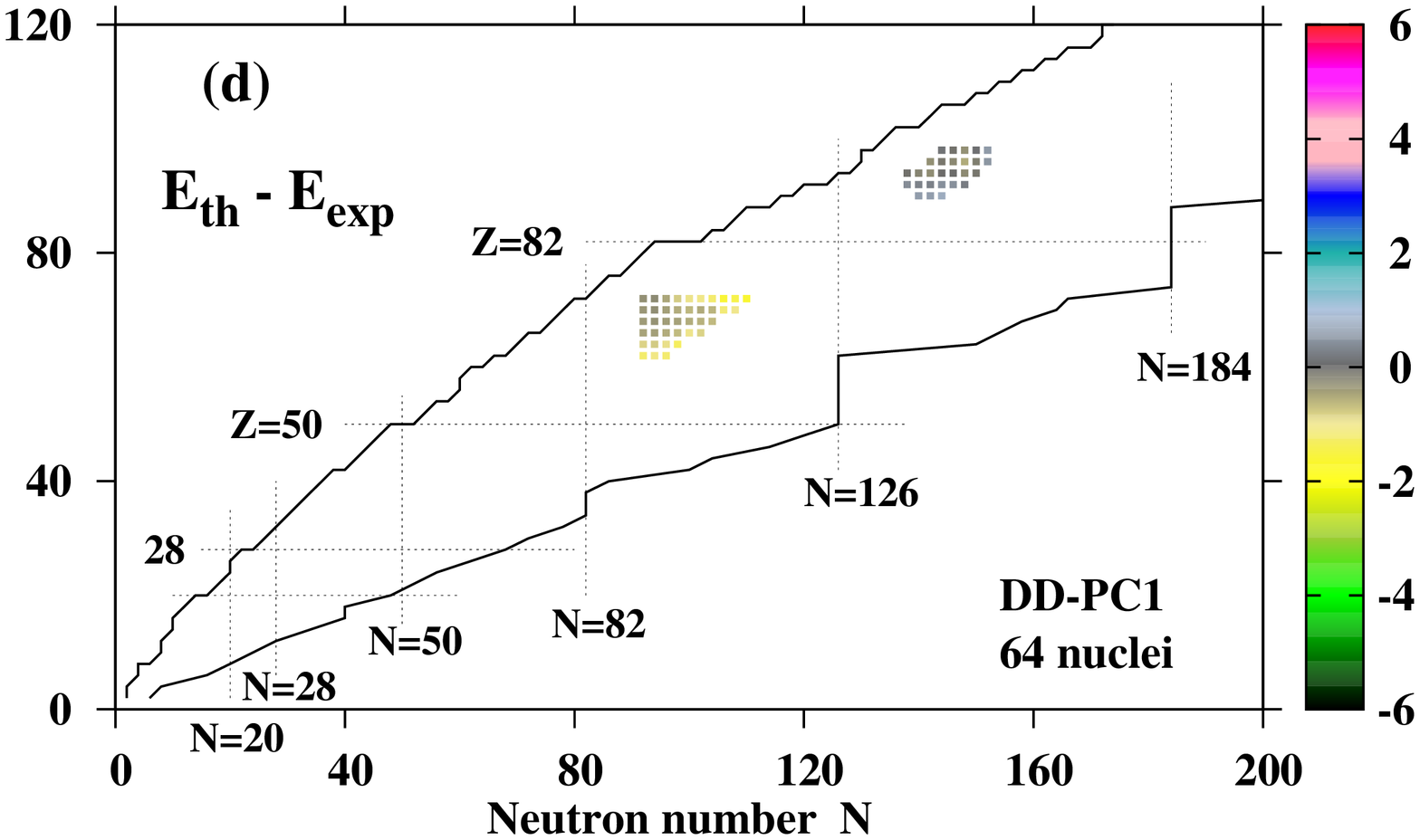}
\caption{(Color online) The nuclei (solid squares), shown in the $(Z,N)$
plane, which were used in the fit of indicated CEDFs. Their total number 
is shown below the functional label. Magic shell closures are shown by 
dashed lines. The colors of the squares show the difference $E_{th}-E_{exp}$
between calculated and experimental binding energies. Two-proton and 
two-neutron drip lines of the indicated functional are shown by solid 
black lines.
}
\label{E_fitted}
\end{figure*}

\begin{figure*}[ht]
\includegraphics[angle=-90,width=8.8cm]{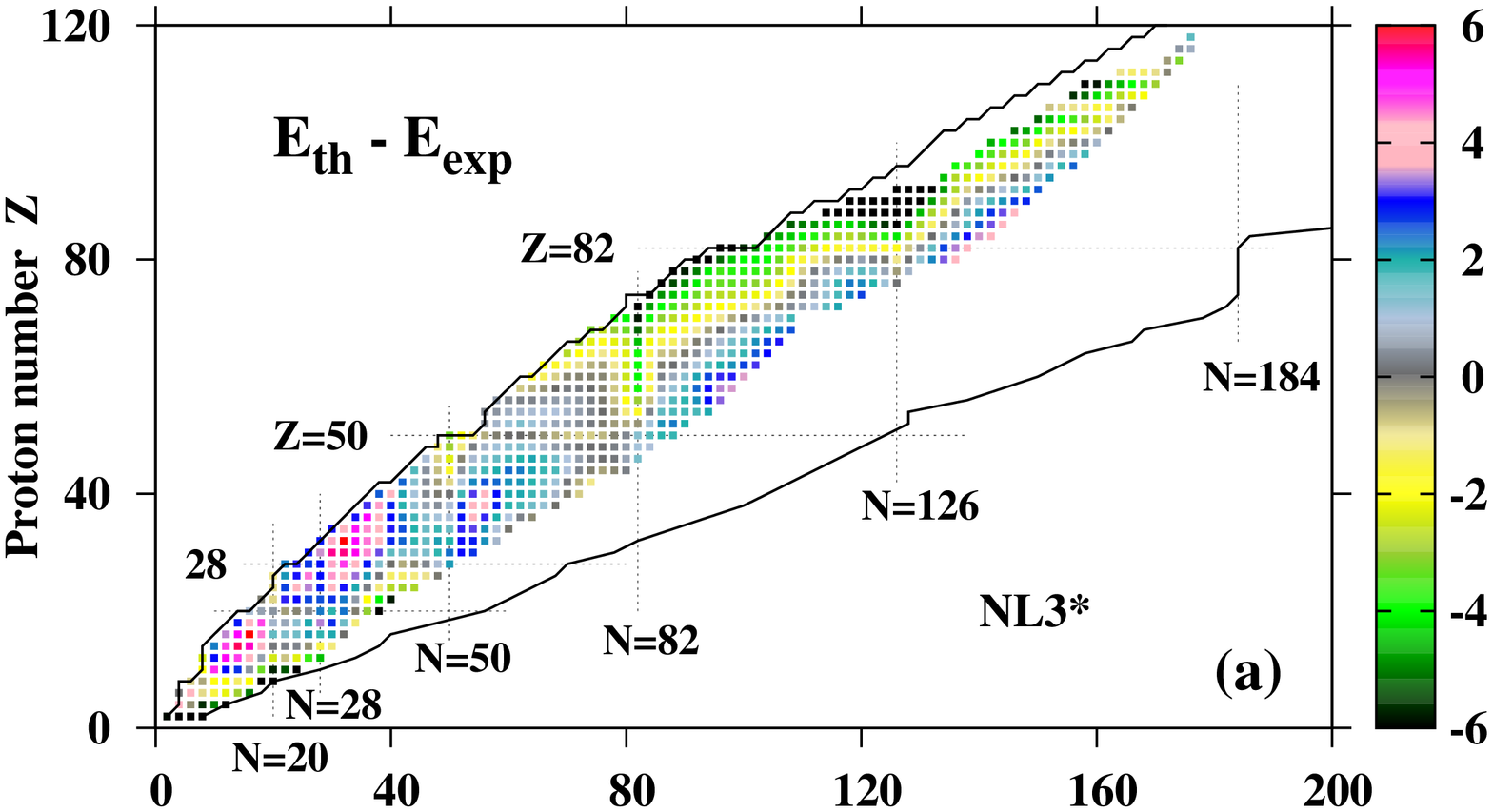}
\includegraphics[angle=-90,width=8.8cm]{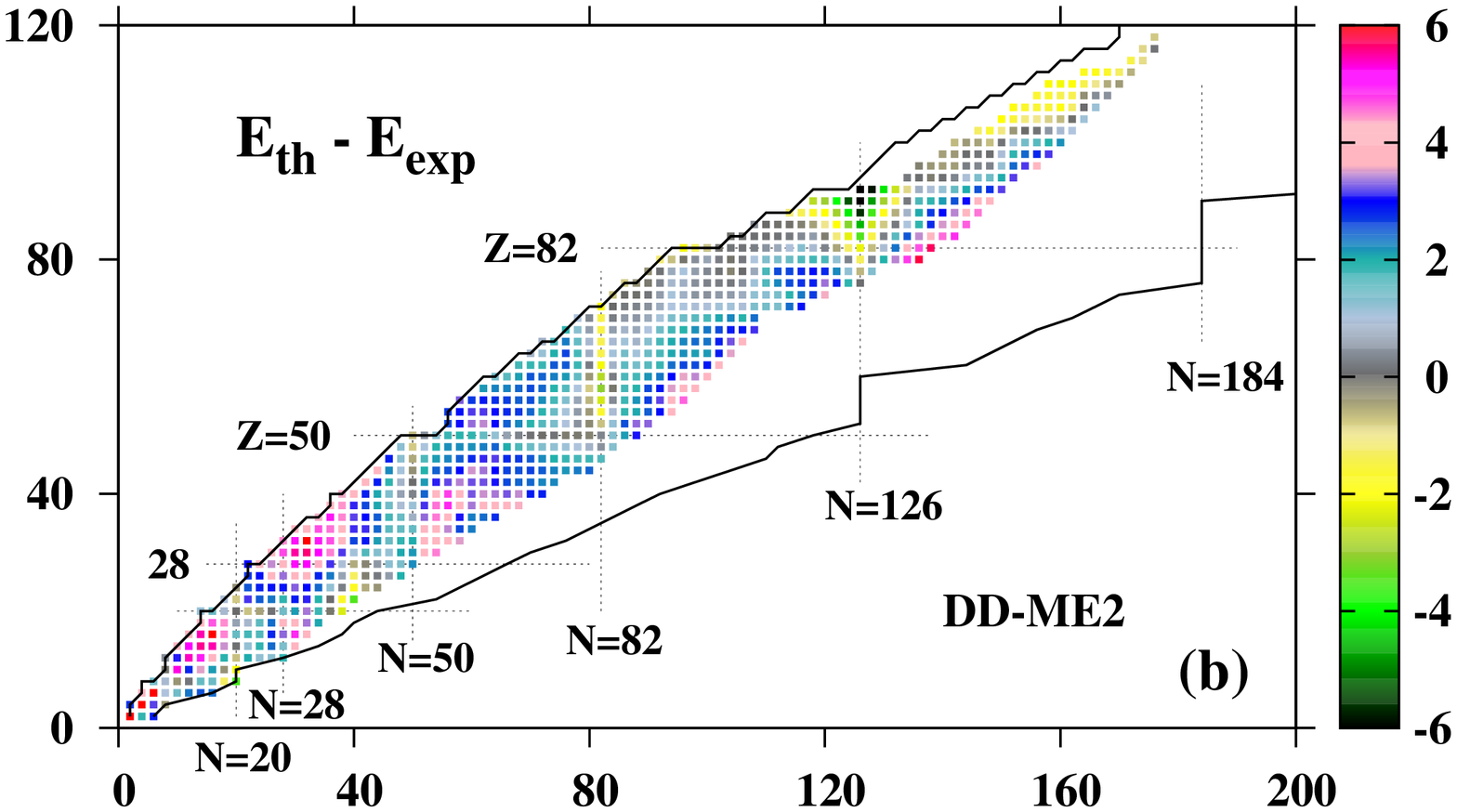}
\includegraphics[angle=-90,width=8.8cm]{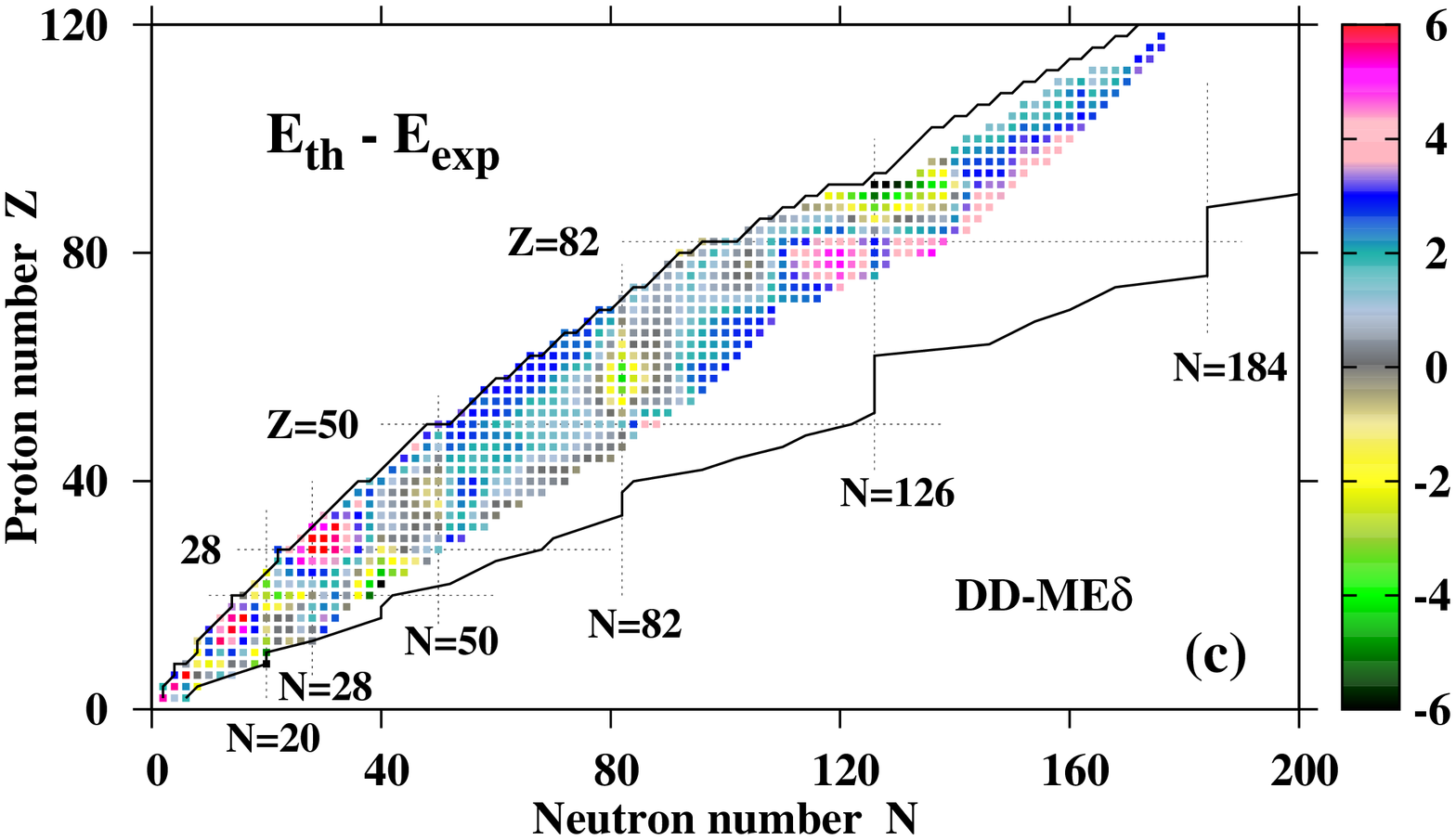}
\includegraphics[angle=-90,width=8.8cm]{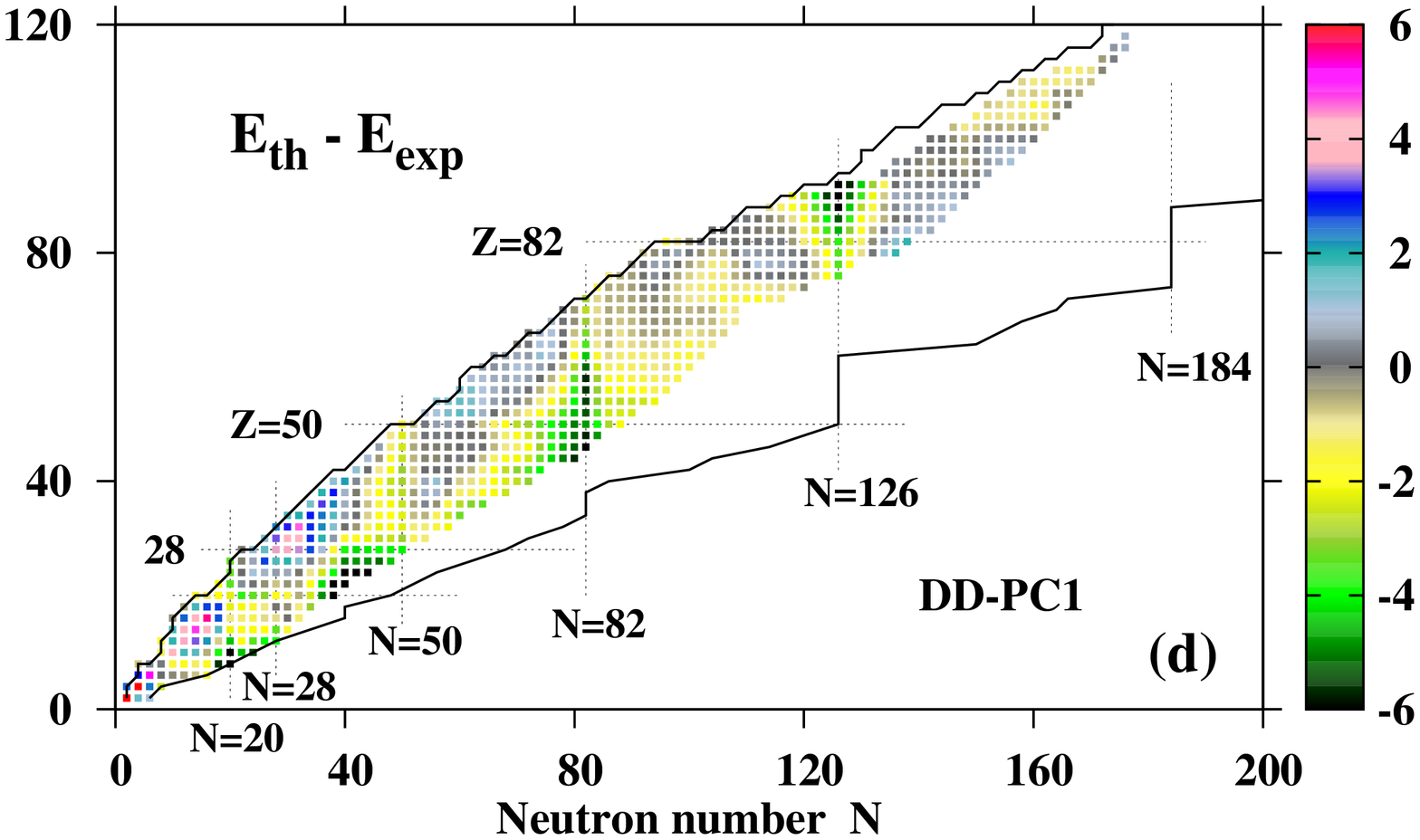}
\caption{(Color online) The differences $E_{th}-E_{exp}$ between 
calculated and experimental binding energies for the indicated
CEDFs. The experimental data are taken from Ref.\ \cite{AME2012} 
and all 835 even-even nuclei, for which measured and estimated 
masses are available, are included. If $E_{th}-E_{exp}<0$, the nucleus 
is more bound in the calculations than in experiment. Two-proton 
and two-neutron drip lines of the indicated functional are shown 
by solid black lines.}
\label{E_th_vs_exp}
\end{figure*}

   As mentioned in the introduction among the functionals for which 
global analysis of experimental binding energies exists only two, 
namely, FSUGold and DD-ME$\delta$ satisfy the majority of the NMP 
constraints. However, CEDFs FSUGold and DD-ME$\delta$ face significant 
problems in the description of finite nuclei. FSUGold is designed for 
neutron star applications in Ref.\ \cite{FSUGold} and it is characterized 
by the largest rms deviations (6.5 MeV) from experiment for binding energies  
among all CEDF's the global performance of which is known \cite{AARR.14}. At 
present, DD-ME$\delta$ is the most microscopic functional among all existing 
CEDFs; it relies on the 
pseudodata from {\it ab initio} calculations to determine density dependence 
of the meson-nucleon vertices so that only 4 parameters are fitted to the 
properties of finite nuclei. Although DD-ME$\delta$ provides quite reasonable 
description of binding energies (see Table \ref{rms-fit-global2}, Fig.\ 
\ref{E_th_vs_exp} below and Refs.\ \cite{AARR.14,AANR.15}), it generates 
unrealistically low inner fission barriers in superheavy elements \cite{AANR.16} 
and fails to reproduce octupole deformed nuclei in actinides \cite{AAR.16}.

 The analysis of Refs.\ \cite{AARR.14,AAR.10,AO.13,AANR.15,AAR.16} clearly 
indicates that the CEDFs NL3*, DD-ME2, PC-PK1 and DD-PC1 represent better 
and well-rounded functionals as compared with FSUGold and DD-ME$\delta$. They 
are able to describe not only ground state properties but also the properties 
of excited states \cite{NL3*,AS.11,AO.13,AAR.10,Meng2013Front.Phys.55,LZZ.14,
PNLV.12,DD-ME2,DD-PC1}. This is despite the fact that first three functionals 
definitely  fail to describe some of the nuclear matter properties (see Table 
\ref{Table-mass-NMP} and Ref.\ \cite{RMF-nm}). It is not clear whether that 
is also a case for DD-PC1 
since it was not analyzed in Ref.\ \cite{RMF-nm}. As a result, one can conclude  
that the functionals, which provide good NMPs, do not necessary well describe 
finite nuclei. Such a possibility has already been mentioned in Ref.\ 
\cite{RMF-nm}.  This is also in line with the results obtained for Skyrme
EDFs \cite{SGSD.13} that the functionals reproducing NMP constraints cannot be 
necessarily expected to reproduce finite nuclei data, to which they were not 
fitted, with very high accuracy. 

 As a consequence, the NMP constraints do not allow to eliminate some 
of the CEDFs from the consideration and in this way to decrease the 
uncertainties in the predictions of binding energies of the neutron-rich 
nuclei and the position of two-neutron drip line;  for a latter see also
Sec.\ VIII in Ref.\ \cite{AARR.14}. Considering substantial 
uncertainties in NMP (see Table \ref{Table-mass-NMP} and Ref.\ \cite{RMF-nm}), 
it is clear that even the combination of their strict enforcement and
the use of large data set on finite nuclei in the fitting protocols of 
new CEDFs will not lead to a substantial lowering or an elimination of the 
uncertainties in the predictions of binding energies of neutron-rich nuclei.

\section{Finite nuclei: the accuracy of the description of the ground 
         state properties and its dependence on fitting protocol}
\label{sec-finite}

  It is clear that the part of the difference in binding energy 
predictions is coming from the use of different data on finite nuclei 
in fitting protocols. For example, the binding energies of the 
``fitted'' nuclei provide the normalization of the energy for the 
functional. Fitting protocols differ substantially (see Table \ref{table-fit} 
and Fig.\ \ref{E_fitted}) and it is important to understand how 
they affect the global results.

  Almost exactly the same fitting protocols exist in the case 
of the NL3* and DD-ME2 functionals which were fitted to the same 12 
spherical nuclei and the same ``empirical'' data on nuclear matter 
properties has been used in the fit \cite{NL3*,DD-ME2}. The only 
difference between them is the fact that 4 and 3 neutron skin 
thicknesses were  used in the fit of NL3* and DD-ME2 CEDFs, 
respectively. However, the impact of this difference is expected
to be very small. Note that contrary to the DD-* functionals, NL3* 
CEDF does not have nonlinearities in the isovector channel. This 
leads to a relatively large values for the symmetry energy $J$ 
and its slope $L_0$ at saturation (see Table \ref{Table-mass-NMP}). 
As a result, the comparison of the calculated and experimental 
binding energies reveals that the DD-ME2 functional has better 
isovector properties than NL3* (Figs.\ \ref{E_th_vs_exp}a and b).

\begin{figure*}[ht]
\includegraphics[angle=-90,width=8.8cm]{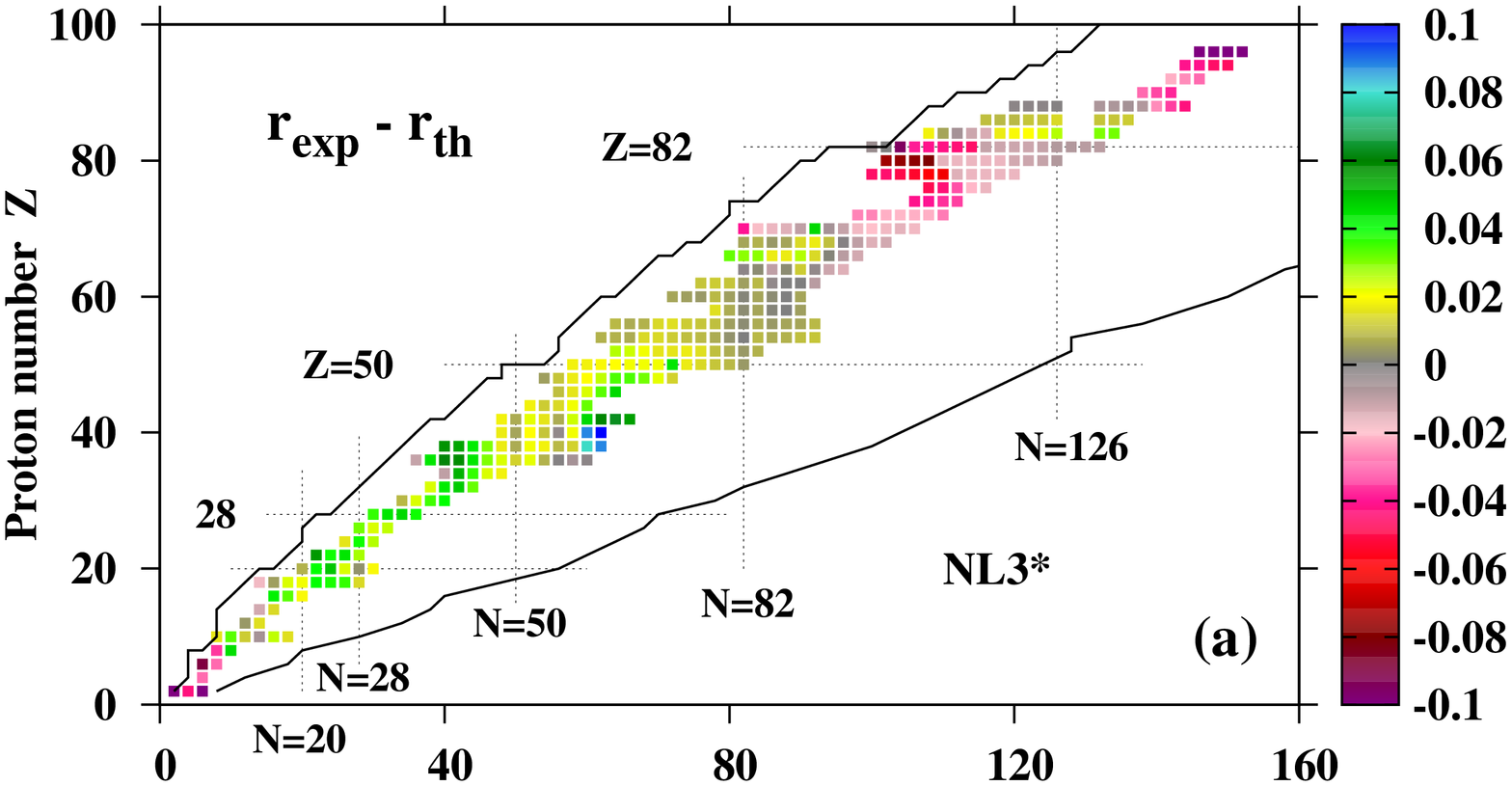}
\includegraphics[angle=-90,width=8.8cm]{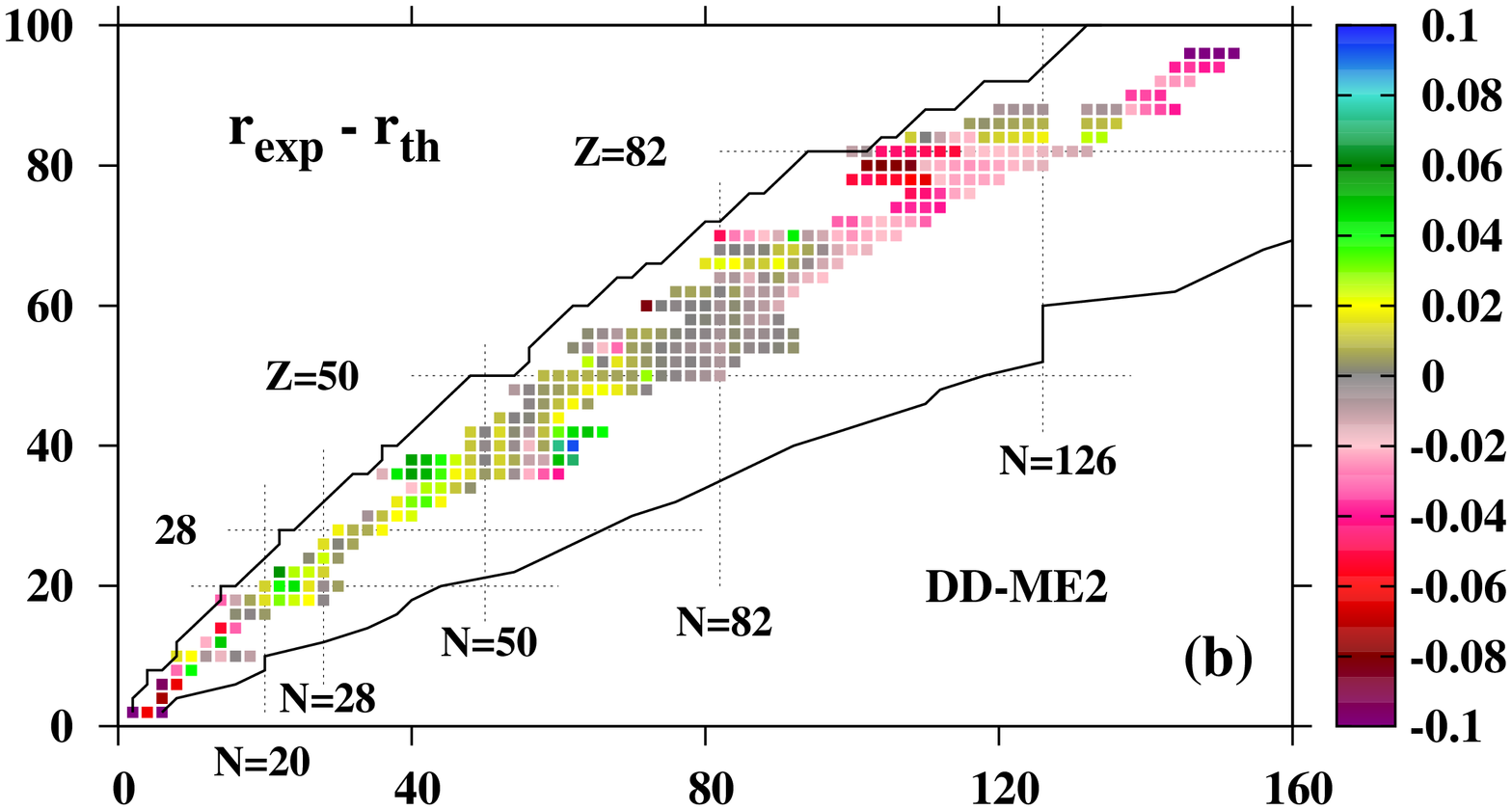}
\includegraphics[angle=-90,width=8.8cm]{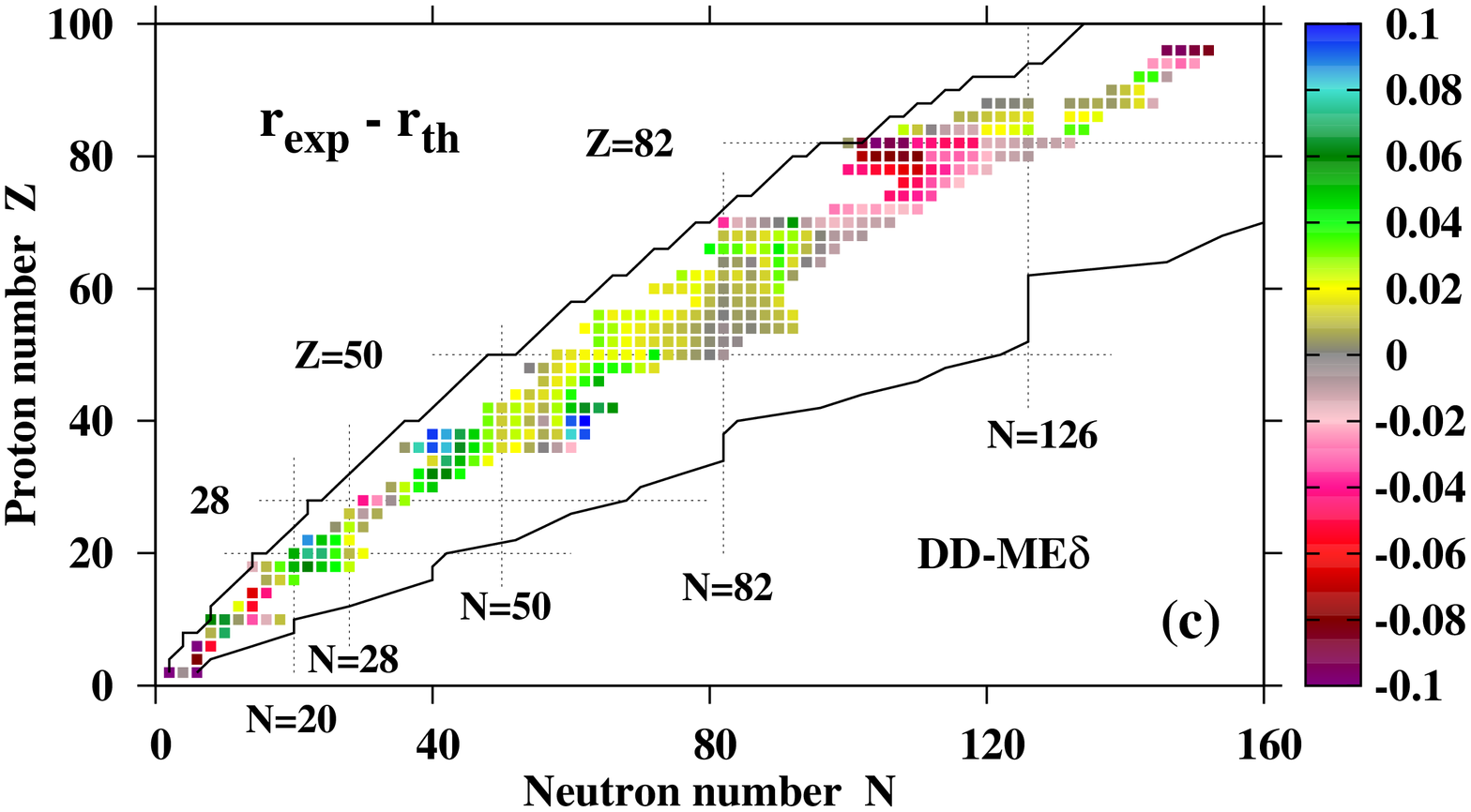}
\includegraphics[angle=-90,width=8.8cm]{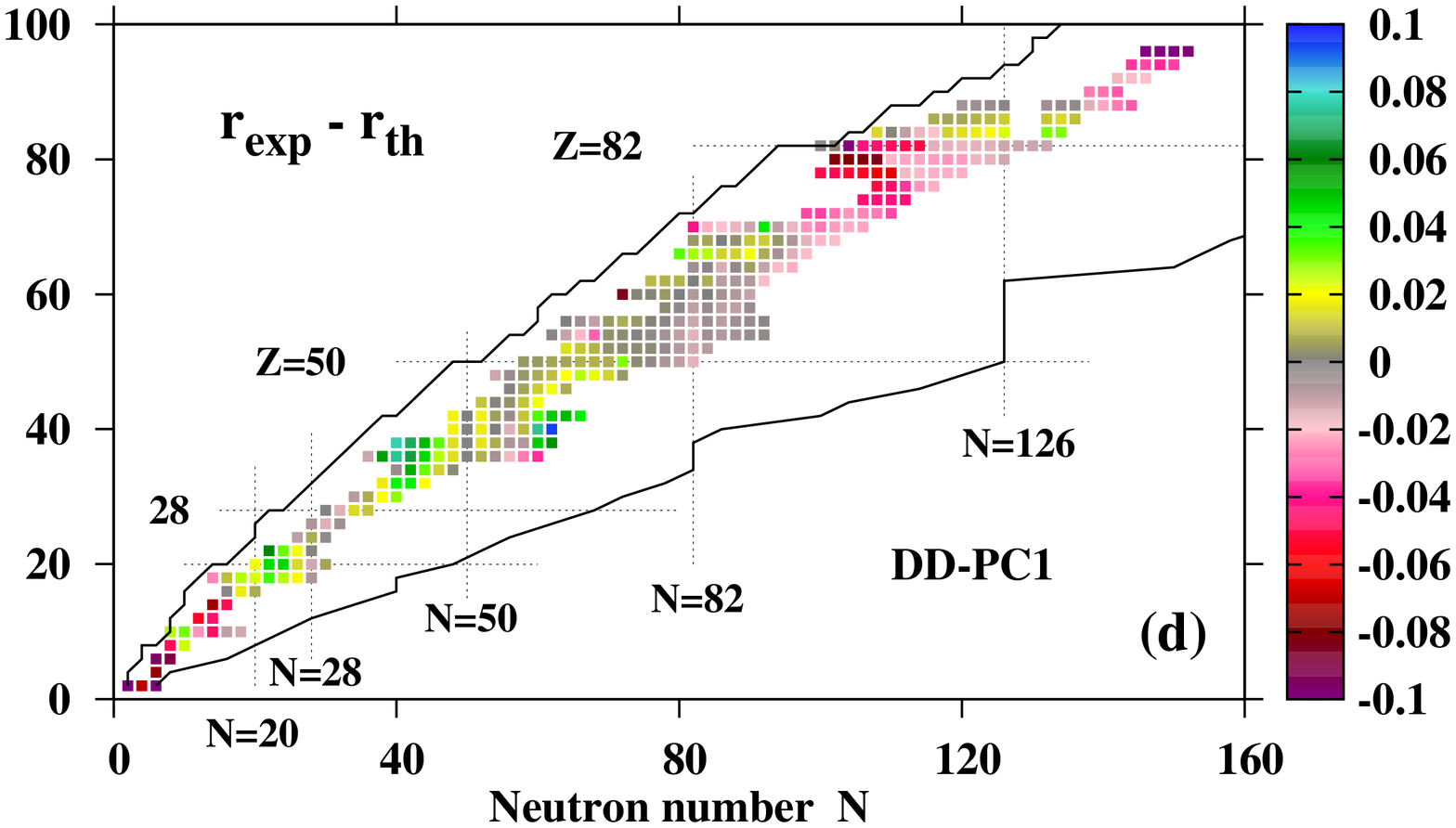}
\caption{(Color online) The difference between measured and calculated
charge radii $r_{ch}$ for indicated  functionals. The experimental 
data are taken from Ref.\ \cite{AM.13}. Two-proton 
and two-neutron drip lines of the indicated functional are shown 
by solid black lines.}
\label{R_ch_th_vs_exp}
\end{figure*}

\begin{figure*}[ht]
\includegraphics[angle=-90,width=8.8cm]{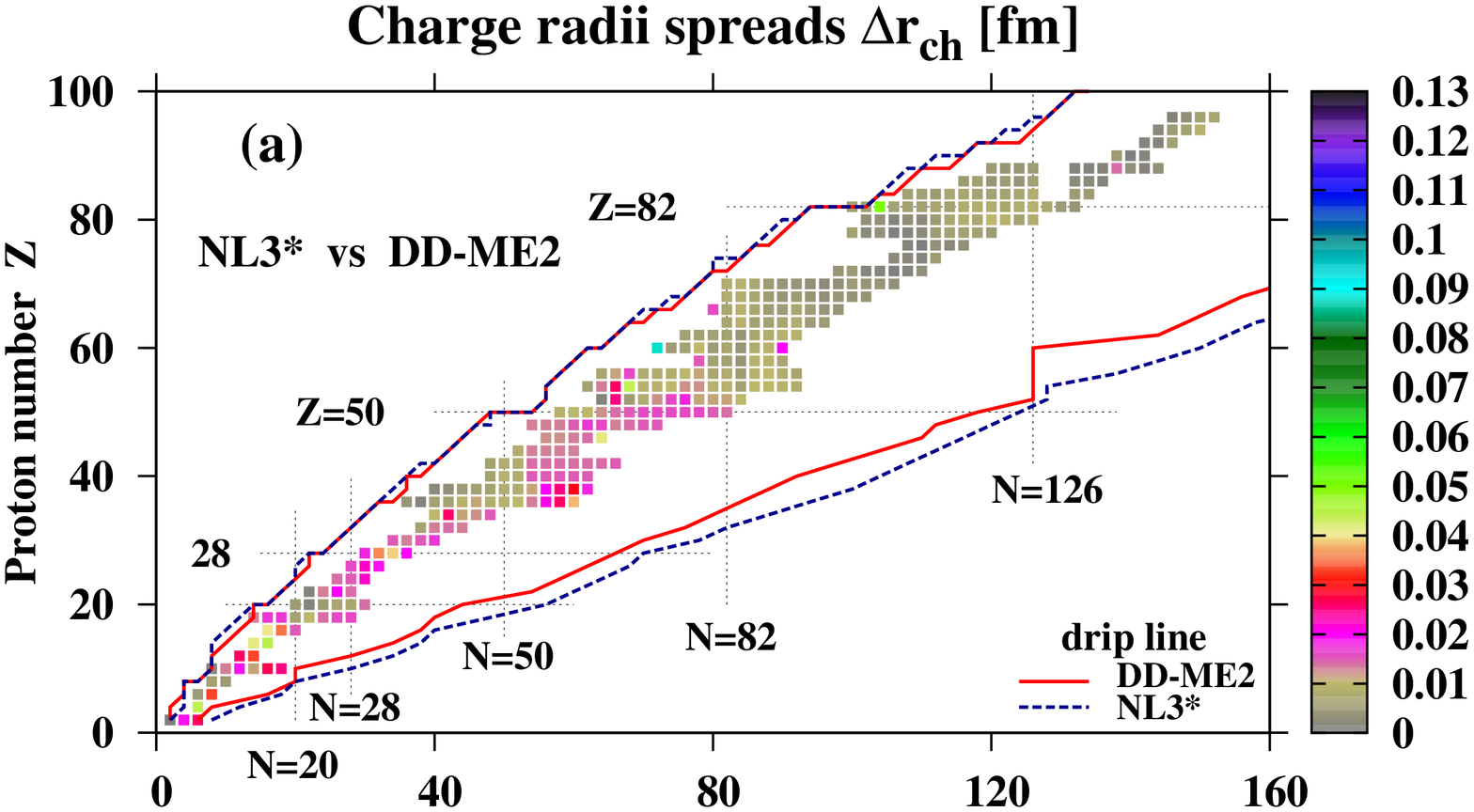}
\includegraphics[angle=-90,width=8.8cm]{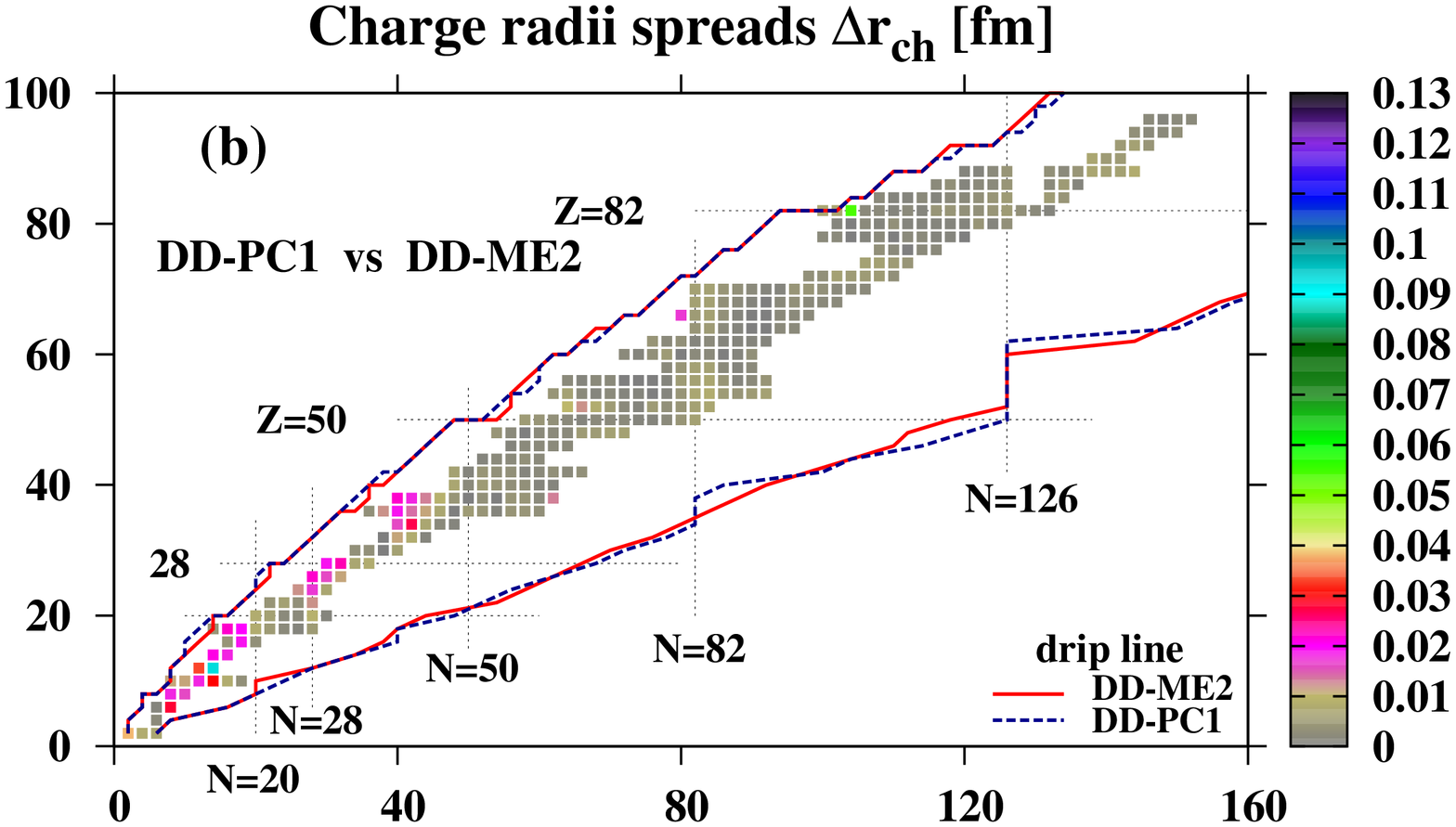}
\includegraphics[angle=-90,width=8.8cm]{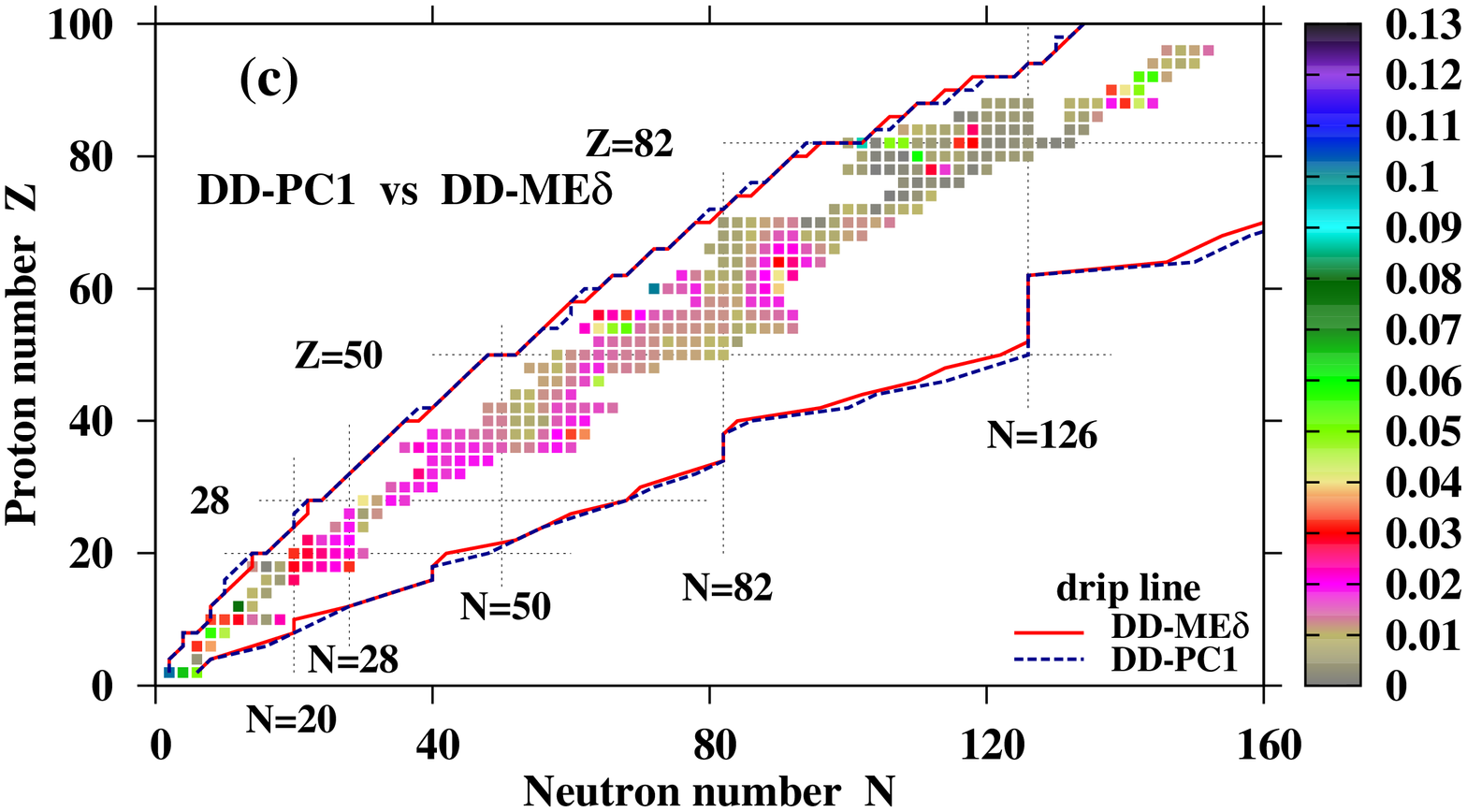}
\includegraphics[angle=-90,width=8.8cm]{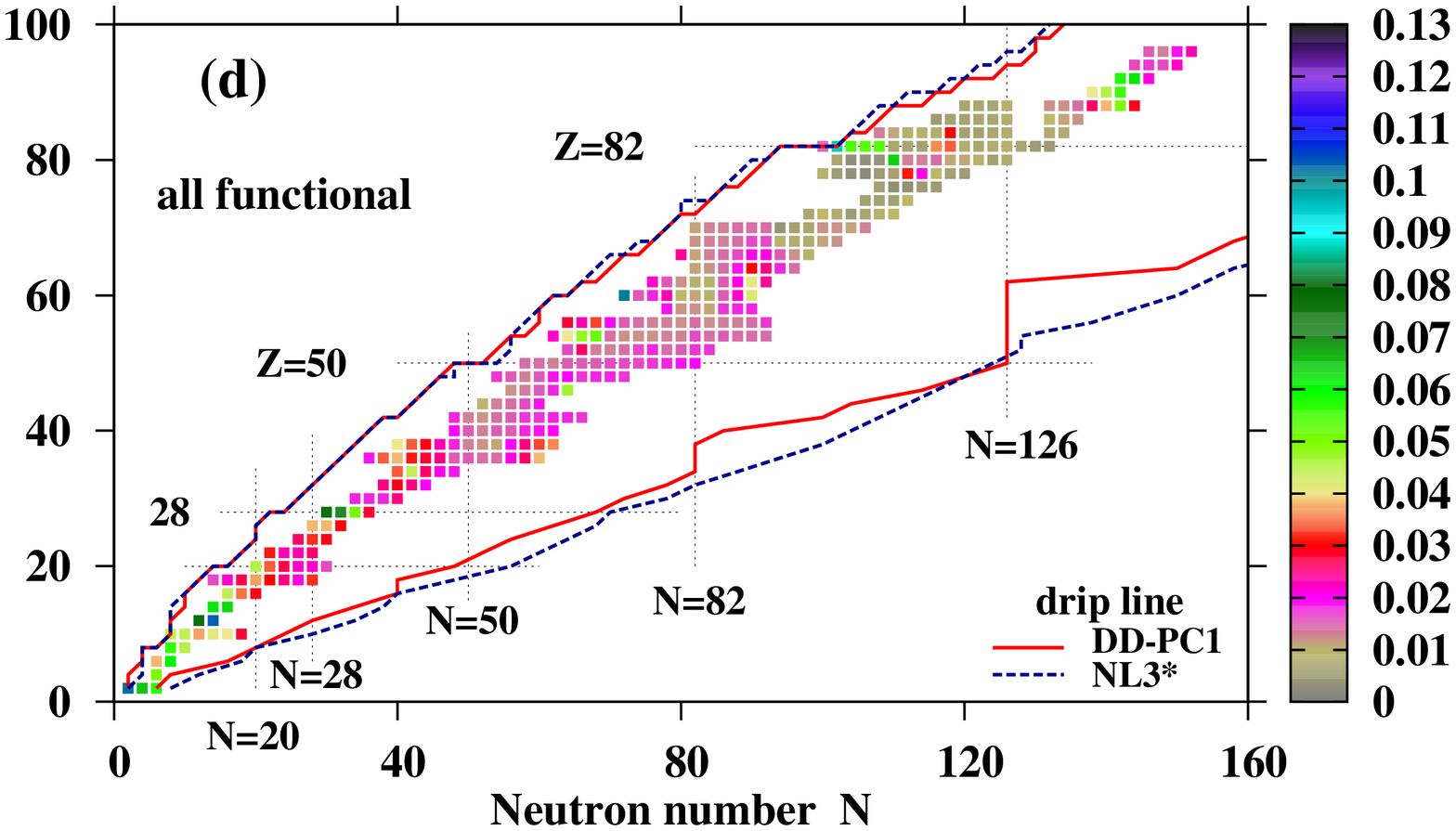}
\caption{(Color online) Charge radii spreads $\Delta r_{ch}(Z,N)$ as a 
function of proton and neutron number. 
$\Delta r_{ch}(Z,N)=|r_{ch}^{max}(Z,N)-r_{ch}^{min}(Z,N)|$,
where $r_{ch}^{max}(Z,N)$ and $r_{ch}^{min}(Z,N)$ are the largest and smallest charge
radii obtained either with indicated pairs of the CEDFs (panels (a)-(c)) or with 
four CEDFs (panel (d)) for the $(Z,N)$ nucleus.}
\label{R_th_vs_th}
\end{figure*}

\begin{figure*}[ht]
\includegraphics[angle=-90,width=11.8cm]{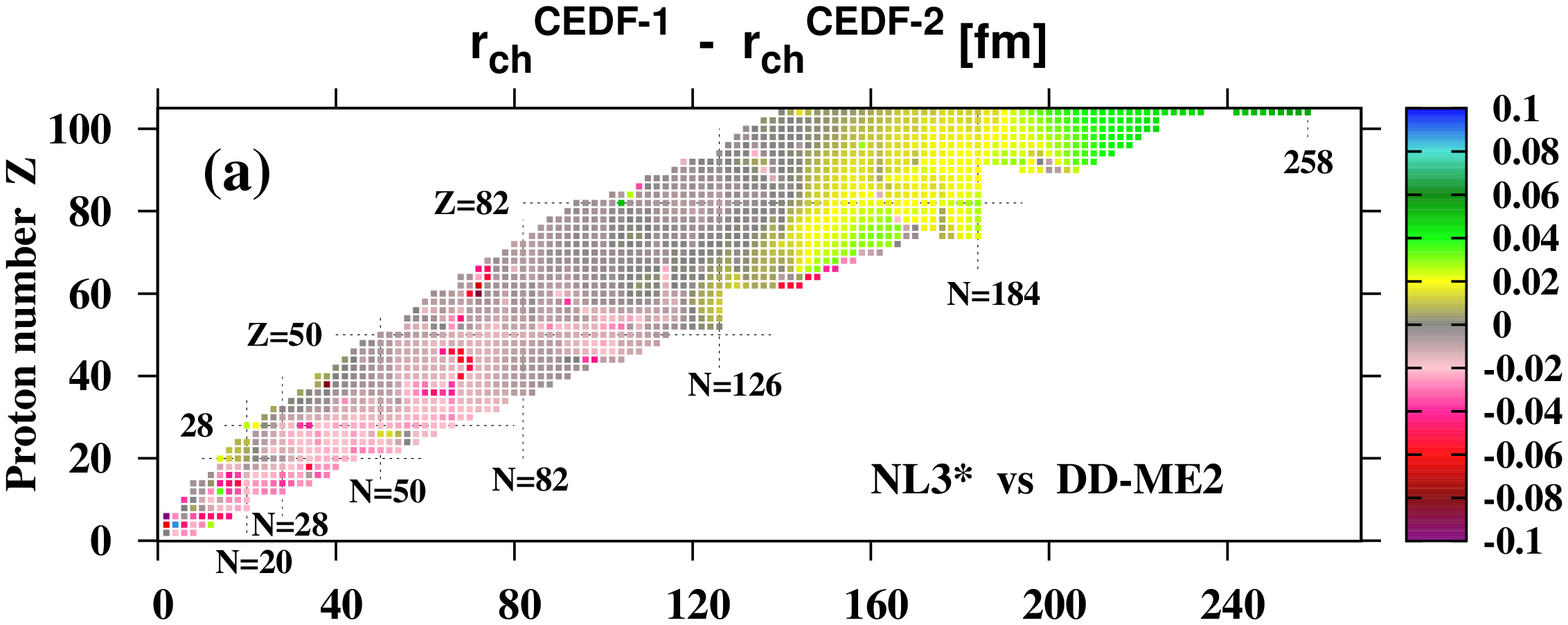}
\includegraphics[angle=-90,width=11.8cm]{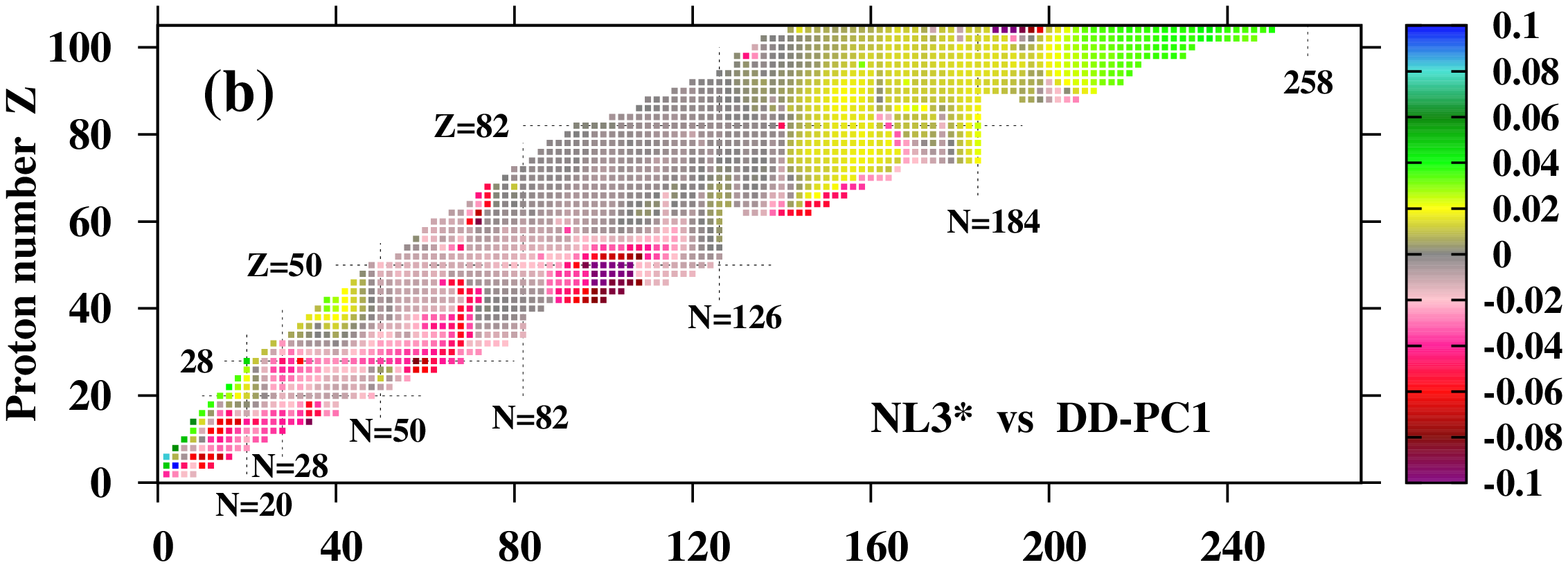}
\includegraphics[angle=-90,width=11.8cm]{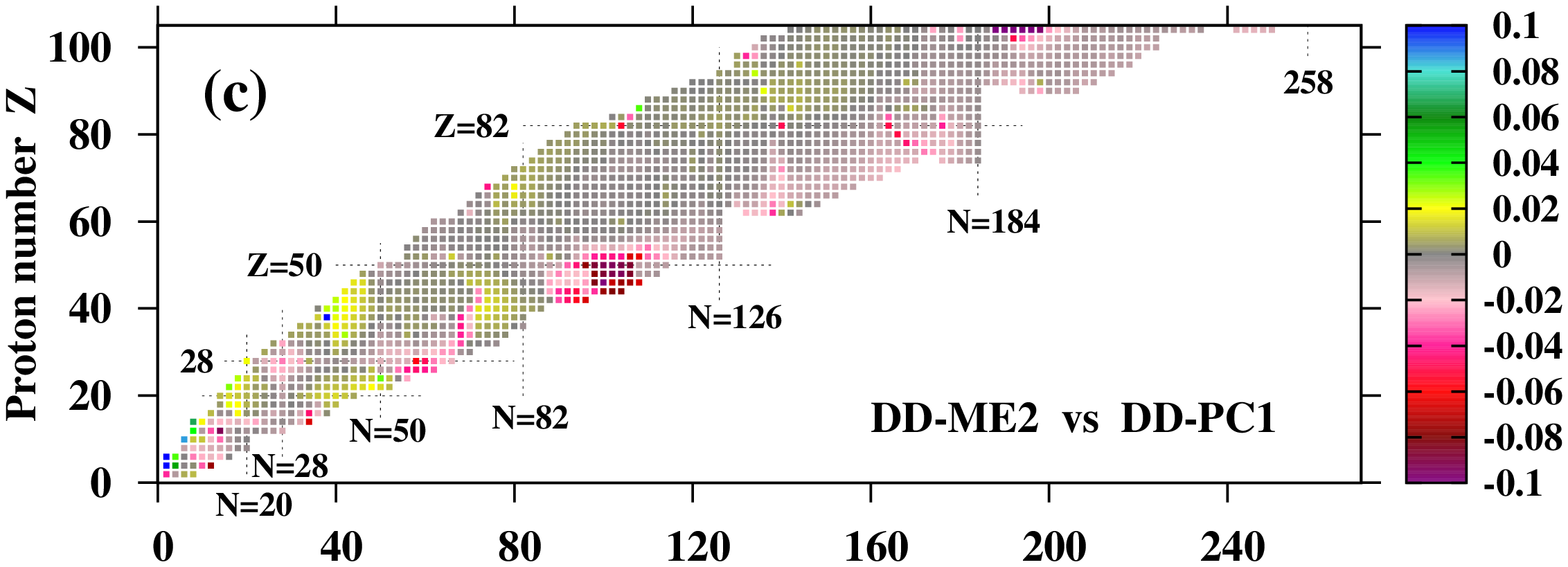}
\includegraphics[angle=-90,width=11.8cm]{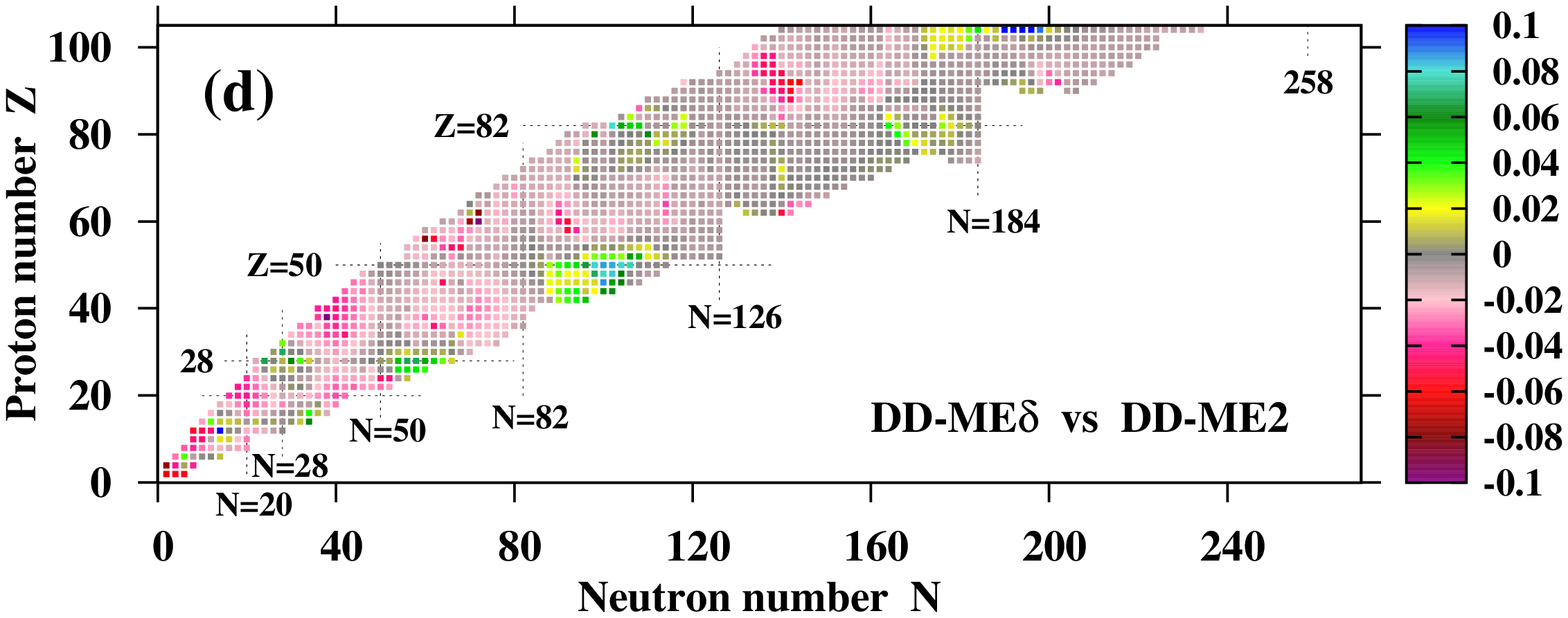}
\caption{(Color online) The difference $r_{ch}^{CEDF-1}(Z,N)-r_{ch}^{CEDF-2}(Z,N)$
in the charge radii predicted by two indicated CEDFs. The second
functional in the label ``CEDF-1 vs CEDF-2'' indicates the reference 
functional. All even-even nuclei between the two-proton and 
two-neutron drip lines are included in the comparison.}
\label{R_ch_th1_vs_th2}
\end{figure*}

  This also leads to somewhat better global reproduction of the 
charge radii in the DD-ME2 functional (see Table \ref{rms-fit-global2} 
and Figs.\ \ref{R_ch_th_vs_exp}a and b). However, apart of few nuclei 
there is basically no difference in the description of experimental
charge radii by NL3* and DD-ME2 for the $Z > 50$ nuclei (Fig.\ 
\ref{R_th_vs_th}a) which suggests that the radii of these medium 
and heavy mass nuclei are not sensitive to non-linearities in the 
isovector channel. On the contrary, the Sn isotopes and lighter 
nuclei show much large sensitivity to the non-linearities in the 
isovector channel. This is an important feature which has to be 
taken into account when considering the use of future data on neutron-rich 
nuclei in the fit of new generation of the functionals. Note that for 
light nuclei beyond mean field effects could be more important than 
for heavy ones and this could be a possible reason for the deterioration 
of the accuracy of the description of the masses and radii in light systems 
as compared with heavy ones. It is also important to mention that despite 
the large similarities in the description of the charge radii in known 
nuclei by the CEDFs NL3* and DD-ME2 (Figs.\ \ref{R_ch_th_vs_exp}a and b
and Fig.\ \ref{R_th_vs_th}a), 
there is a substantial increase of the differences in the predictions of 
the charge radii by these two CEDFs for the part of the nuclear chart 
roughly characterized by particle numbers $Z > 70$ and $N >140$ (Fig.\ 
\ref{R_ch_th1_vs_th2}a).

 The DD-* CEDFs show better isovector properties in the description
of binding energies as compared with the NL3* one (Fig.\ \ref{E_th_vs_exp}). 
However, there are some important differences between the meson-exchange 
functionals (DD-ME2 and DD-ME$\delta$) and point-coupling functional 
DD-PC1. They are the consequences of both the selection of the finite
nuclei for the fitting protocol (Table \ref{table-fit} and Fig.\ \ref{E_fitted}) 
and the differences in underlying physics (meson exchange versus point 
coupling). Meson-exchange functionals are fitted to the spherical nuclei.
As a result, deformed nuclei are typically underbound in these functionals
(Fig.\ \ref{E_th_vs_exp}b and c). In contrast, DD-PC1 CEDF is fitted 
to the deformed nuclei and it reproduces the binding energies of such 
nuclei rather well especially in the rare-earth region and actinides 
(Fig.\ \ref{E_fitted}d). However, this functional tends to overbind 
spherical nuclei.

  The differences in the underlying physics show themselves in different 
isovector properties of point coupling and meson exchange functionals. For 
example, similar description of experimental binding energies of neutron-deficient 
$72<Z<96$ nuclei is achieved in all DD-* functionals (Figs.\ \ref{E_th_vs_exp}b, 
c and d). However,  neutron-rich $72<Z<96$ nuclei are underbound in the RHB 
calculations with CEDFs DD-ME2 and DD-ME$\delta$ (Figs.\ \ref{E_th_vs_exp}b and 
c) and overbound in the ones with DD-PC1 (Fig.\ \ref{E_th_vs_exp}d). Similar
situation exists also for lighter nuclei in CEDF DD-PC1; neutron rich nuclei 
are overbound in the calculations while neutron deficient ones are either 
underbound or close to experiment. Thus, for DD-PC1 functional with increasing 
neutron number the binding energies increase faster in the calculations than in 
experiment across whole nuclear chart. The same trend for rate of binding 
energy changes is seen also in the CEDFs DD-ME2 and DD-ME$\delta$ for light nuclei
(Figs.\ \ref{E_th_vs_exp}b and c); this is contrary to the situation in the
heavy nuclei.  Thus, the isovector dependence of binding energies is different 
in light and heavy nuclei in meson-exchange and point coupling functionals. Note 
that above discussed general trends are somewhat disturbed by local differences 
which emerge from the differences in the underlying shell structure.

  Fig.\ \ref{R_ch_th_vs_exp} and Table \ref{rms-fit-global2} show the accuracy 
of the description of charge radii by density dependent functionals. One can 
see that CEDFs DD-ME2 and DD-PC1 provide comparable accuracy of the description 
of charge radii. On the other hand, the DD-ME$\delta$ functional provides worst 
description of the radii among considered functionals. Fig.\ \ref{R_th_vs_th}b 
shows that apart of some light nuclei the DD-ME2 and DD-PC1 functionals provide 
almost the same description of charge radii across the nuclear chart. This is 
despite the fact that DD-PC1 functional has been defined without experimental data 
on charge radii in Ref.\ \cite{DD-PC1}. Figs.\ \ref{R_th_vs_th}c and d show 
that the absolute majority of the spreads in charge radii for a set of four 
functionals is coming from the DD-ME$\delta$ functional; the next contributor to 
these spreads is the CEDF NL3* (Fig.\ \ref{R_th_vs_th}).

 However, global predictions for the charge radii are similar for density 
dependent functionals. Indeed, there are no global differences in the 
DD-ME2/DD-PC1 (Fig.\ \ref{R_ch_th1_vs_th2}c) and DD-ME$\delta$/DD-ME2 
(Fig.\ \ref{R_ch_th1_vs_th2}d) pairs of CEDFs similar to the ones observed 
in the NL3*/DD-ME2 (Fig.\ \ref{R_ch_th1_vs_th2}a) and NL3*/DD-PC1  
(Fig.\ \ref{R_ch_th1_vs_th2}b) pairs for the part of the nuclear chart roughly 
characterized by particle numbers 
$Z > 70$ and $N >140$. However, the local differences emerging from the underlying 
shell structure clearly exist. For example, substantial differences in charge radii 
seen at $Z\sim 90, N\sim 134$ for the DD-ME$\delta$/DD-ME2 pair of the functionals 
(Fig.\ \ref{R_ch_th1_vs_th2}d) are due to inability of the CEDF DD-ME$\delta$ to 
describe octupole deformed nuclei in the actinides \cite{AAR.16}.

\section{General observations}
\label{sec-general}
 
  Based on the present analysis one could make the following
observations:

\begin{itemize}

\item
  Fig.\ \ref{E_spreads} shows that fastest increase of the 
differences in predicted binding energies of two functionals 
takes place in the direction which is perpendicular to the 
gray band of similar energies (or beta-stability line). 
These differences are due to (i) different isovector 
properties of these functionals and (ii) the differences
in the selection of the input for the fitting protocols
of these functionals.

 The differences in the NL3*/DD-ME2 pair of the functionals
are mostly due to different isovector properties of compared 
functionals; they form smooth trends with almost no local 
fluctuations for the differences of both the binding energies (Fig.\ 
\ref{E_spreads}c) and charge radii (Fig.\ \ref{R_ch_th1_vs_th2}a).
This is a consequence of the same fitting protocol used for both
functionals which leads to a global similarity of their underlying 
single-particle structure.

  The differences in the input of the fitting protocols lead
to local deviations from smooth trends which become especially 
visible in the case of the NL3*/DD-PC1 and DD-ME$\delta$/DD-ME2 
pairs of the functionals (see Figs.\ \ref{E_spreads}a and b for 
the differences of the binding energies and Figs.\ \ref{R_ch_th1_vs_th2}b 
and d for the differences of the charge radii). They are due to local 
differences in the underlying single-particle structure of the 
compared functionals.

  Thus, the selection of fitting protocol and in particular the 
selection of the information on finite nuclei has a direct 
influence and creates an imprint on the global performance of 
the CEDF.

\item
 An important question is how much and which type of the data on 
finite nuclei are essential in the fitting protocols. Table \ref{rms-fit-global2} 
suggests that overdefined fitting protocols (such as DD-ME$\delta$ with 
161 binding energies and 86 charge radii) do not offer any advantages as 
compared with the protocols which contain much less data since 
$\Delta E^{fit}_{rms} \approx \Delta E^{global}_{rms}$ and 
$\Delta(r_{ch})^{fit}_{rms} \approx \Delta(r_{ch})^{global}_{rms}$ for this functional. 
Note that DD-ME$\delta$ is only the functional with such properties; for all 
other functionals $\Delta E^{fit}_{rms} < \Delta E^{global}_{rms}$ and 
$\Delta(r_{ch})^{fit}_{rms} < \Delta(r_{ch})^{global}_{rms}$ (see Table 
\ref{rms-fit-global2}). The DD-PC1 CEDF \cite{DD-PC1} is an example of the 
functional which achieves good global description of charge radii (Table 
\ref{rms-fit-global2}) without any experimental data on charge radii in the 
fitting protocol (Table \ref{table-fit}).  Curiously enough the addition of 
experimental data on charge radii to the fitting protocol of DD-PC1 could lead 
to the deterioration of the accuracy of the description of charge radii 
since $\Delta(r_{ch})^{fit}_{rms} > \Delta(r_{ch})^{global}_{rms}$ (Table 
\ref{rms-fit-global2}). It turns out that the inclusion of very few 
but carefully selected experimental charge radii (like in the CEDF DD-ME2 
with 9 charge radii) leads to a further but moderate increase of an accuracy
of the global description of charge radii. The analysis of Table 
\ref{rms-fit-global2} also suggests that less than 100 data points
on binding energies is sufficient for fitting protocols of a current 
generation of the CEDFs not aimed at the ``mass table'' quality description
of binding energies.

\item
 The variations of the differences in the predictions of the binding 
energies and the differences in the predictions of the charge radii 
of two functionals with proton and neutron numbers shown in Figs.\ 
\ref{E_spreads} and \ref{R_ch_th1_vs_th2} are not globally correlated.

\item
  It is necessary to recognize that the binding energies are affected 
by the effects beyond mean field \cite{BBH.06,GCP.09,LLLYM.15} which
are not included in the current calculations. It was shown in Ref.\ 
\cite{LLLYM.15} for the PC-PK1 functional that the inclusion of dynamic 
correlation energies (DCE) leads to a reduction of the rms-deviations 
for binding energies of 575 known even-even nuclei from 2.52 MeV (at 
the mean field level) to 1.14 MeV. DCE provides an additional binding 
and vary mostly in the region of $2.0-3.5$ MeV. It is expected that DCE 
will depend relatively weakly on the underlying functional (Ref.\ 
\cite{LLLYM.15}). Thus, the accounting of DCE will not remove existing 
differences between the functionals seen in Figs.\ \ref{DE_spread_total}, 
\ref{FRIB-data-constraint}, \ref{Yb-with-PCPK1} and \ref{E_spreads}.

  On the other hand, the rms-deviations between  experimental and 
calculated binding energies will be reduced for the DD-ME2 and 
DD-ME$\delta$ CEDFs when DCE are taken into account since these functionals 
typically underbind nuclei (see Figs.\ \ref{E_th_vs_exp}b and c).
However,  they will be increased for DD-PC1 since it provides good 
description of the binding energies of rare-earth nuclei and actinides at 
the mean field level (Fig.\ \ref{E_th_vs_exp}d). The inclusion of DCE will 
have probably overall neutral effect on the NL3* CEDF since it will improve the 
description of the binding energies of the $Z\leq 50$ nuclei but will 
lead to the decrease of the accuracy of the description of heavier nuclei 
(Fig.\ \ref{E_th_vs_exp}a).

\item
  It is well known that high-density behaviour of EOS has little 
influence on the description of low-energy nuclear structure data 
\cite{DD-PC1}. For the first time our studies confirm this fact on a 
global scale in the framework of CDFT. Indeed, the functionals which 
have stiff equation of state (NL3*, DD-ME2 and PC-PK1) are still accurate 
in the description of the ground state properties. The current analysis 
clearly indicates that apart of the NL3* functional, which does not 
have a good reproduction of the isospin trends for the binding energies, 
the remaining functionals are quite comparable at the mean field level. 
Considering similar (as compared with experiment) initial starting mean 
field solutions for CEDFs DD-ME2, DD-ME$\delta$ and PC-PK1 (see Fig.\ 
\ref{E_th_vs_exp} and Fig. 3 in Ref.\ \cite{LLLYM.15}) and weak dependence 
of dynamic correlation energies on the functional, it is reasonable to 
expect that CEDFs DD-ME2 and DD-ME$\delta$ will have similar to PC-PK1 
rms-deviations for binding energies ($\Delta E_{rms}^{global} \sim 1.14$ MeV 
\cite{LLLYM.15}) when DCEs are included. Thus, the functionals with 
$J\sim 32$ MeV and $L_0\sim 50$ MeV (DD-ME2 and DD-ME$\delta$) and 
$J=35.6$ MeV and $L_0=113$ MeV (PC-PK1) provide quite comparable global 
description of the binding energies. For the first time in CDFT, this
 confirms on a global scale earlier observations that nuclear binding 
energies represent poor isovector indicators \cite{NRSV.14}.

  The slope of symmetry energy $L_0$, which is good isovector indicator, 
is not well defined at present (see Table \ref{Table-mass-NMP} and Ref.\ 
\cite{RMF-nm}). However, it strongly correlates with the size of neutron 
skin thickness so that CEDFs NL3* and PC-PK1 predict larger neutron skin 
than DD-* functionals (see Sec.\ X in Ref.\ \cite{AARR.14}). At present, 
the uncertainties in experimental definition of neutron skin still exist. 
However, there is a hope that the PREX-II experiment aimed at the 
measurement of neutron radii in $^{208}$Pb will put a stricter constraint 
on the density dependence of the symmetry energy (the $L_0$ parameter) 
(Ref.\ \cite{HKM.14}).

\item 
 A common trend of the discrepancies between calculated and experimental 
charge radii is clearly seen for all functionals in Fig.\ \ref{R_ch_th_vs_exp}. 
The RHB calculations typically underestimate the radii in the $Z\leq 50$ nuclei,
rather well reproduce them in the rare-earth region and overestimate
them in the actinides. This trend is based on the consideration which 
excludes neutron deficient nuclei in the lead and krypton regions; 
they are  characterized by shape coexistence which cannot be described 
at the mean field level (see Sec.\ X of Ref.\ \cite{AARR.14} for a discussion 
of these nuclei). It remains to be seen to which extent this trend is due 
to the use of the equation
\begin{eqnarray}
r_{ch}=\sqrt{<r^2>_p+0.64}\,\,\,fm,
\end{eqnarray}
for charge radii. The factor 0.64 accounts for the finite-size effects of the proton.
This equation is used in the CDFT calculations \cite{Ring1996_PPNP37-193,GTM.05} 
but it ignores small contributions to the charge radius originating from the 
electric neutron form factor and electromagnetic spin-orbit coupling \cite{BFHN.72,NS.87} 
as well as the corrections due to the center-of-mass motion.

\end{itemize}

\section{CONCLUSIONS}
\label{concl}

 The question of how strictly nuclear matter constraints have to be imposed
and which values have to be used for the definition of covariant energy density 
functionals still remains not fully answered. Definitely, the equation 
of state relating pressure, energy density, and temperature at a given 
particle number density is essential for modeling neutron stars, core-collapse 
supernovae, mergers of neutron stars and the processes (such as nucleosynthesis) 
taking places in these environments. However, there are substantial 
experimental/empirical/model uncertainties in the definition of the 
NMP constraints.

  In addition, the properties of finite nuclei are defined by the 
underlying shell structure which depends sensitively on the single-particle 
features \cite{BHP.03,VALR.05,AANR.15}. As a consequence, we are facing the 
situation in which the functionals which are coming close to satisfying all NMP 
constraints perform quite poorly in the description of finite nuclei. This
was exemplified by the FSUGold and DD-ME$\delta$ functionals. The former
provides the worst rms-deviations in global description of binding energies
\cite{RA.11, AARR.14}, while the latter fails to reproduce octupole deformed 
actinides \cite{AANR.15} and predicts too low fission barriers in superheavy
nuclei \cite{AANR.16} so that their existence could be questioned. On the other 
hand, the functionals which fail to reproduce the NMP constraints suggested in Ref.\ 
\cite{RMF-nm} such as NL3* and PC-PK1 are able to reproduce reasonably well the 
ground state properties of finite nuclei, such as binding energies and charge radii, 
fission barriers \cite{AAR.10,AAR.12,LZZ.14}, rotating nuclei 
\cite{AO.13,Meng2013Front.Phys.55} and the energies of the single-particle states 
in spherical \cite{LA.11,AL.15} and deformed nuclei \cite{AS.11,DABRS.15}.

  The correlations between global description of the binding energies and 
nuclear matter properties of the underlying functionals have been discussed 
based on the results of recent assessment of global performance of covariant 
energy density functionals presented in Refs.\ \cite{AARR.13,AARR.14,AARR.15}.
It was concluded that the strict enforcement of the limits on the nuclear matter 
properties defined in Ref.\ \cite{RMF-nm} will not necessary (i) lead to the 
functionals with good description of binding energies or other ground and
excited state properties or (ii) substantially reduce the uncertainties in the 
description of neutron-rich systems. This is very likely due to to the mismatch 
of phenomenological content, existing in all modern functionals, related to 
nuclear matter physics and the physics of finite nuclei; the latter being 
strongly affected by underlying shell effects.

\section{ACKNOWLEDGMENTS}

 This material is based upon work supported by the U.S. Department of Energy, 
Office of Science, Office of Nuclear Physics under Award Number DE-SC0013037. We 
would like to express our deep graditude to D. Ray and P. Ring, who contributed 
to Ref.\ \cite{AARR.14} the numerical results of which were used in this 
manuscript.

\bibliography{references13}

\begin{thebibliography}{58}%
\makeatletter
\providecommand \@ifxundefined [1]{%
 \@ifx{#1\undefined}
}%
\providecommand \@ifnum [1]{%
 \ifnum #1\expandafter \@firstoftwo
 \else \expandafter \@secondoftwo
 \fi
}%
\providecommand \@ifx [1]{%
 \ifx #1\expandafter \@firstoftwo
 \else \expandafter \@secondoftwo
 \fi
}%
\providecommand \natexlab [1]{#1}%
\providecommand \enquote  [1]{``#1''}%
\providecommand \bibnamefont  [1]{#1}%
\providecommand \bibfnamefont [1]{#1}%
\providecommand \citenamefont [1]{#1}%
\providecommand \href@noop [0]{\@secondoftwo}%
\providecommand \href [0]{\begingroup \@sanitize@url \@href}%
\providecommand \@href[1]{\@@startlink{#1}\@@href}%
\providecommand \@@href[1]{\endgroup#1\@@endlink}%
\providecommand \@sanitize@url [0]{\catcode `\\12\catcode `\$12\catcode
  `\&12\catcode `\#12\catcode `\^12\catcode `\_12\catcode `\%12\relax}%
\providecommand \@@startlink[1]{}%
\providecommand \@@endlink[0]{}%
\providecommand \url  [0]{\begingroup\@sanitize@url \@url }%
\providecommand \@url [1]{\endgroup\@href {#1}{\urlprefix }}%
\providecommand \urlprefix  [0]{URL }%
\providecommand \Eprint [0]{\href }%
\providecommand \doibase [0]{http://dx.doi.org/}%
\providecommand \selectlanguage [0]{\@gobble}%
\providecommand \bibinfo  [0]{\@secondoftwo}%
\providecommand \bibfield  [0]{\@secondoftwo}%
\providecommand \translation [1]{[#1]}%
\providecommand \BibitemOpen [0]{}%
\providecommand \bibitemStop [0]{}%
\providecommand \bibitemNoStop [0]{.\EOS\space}%
\providecommand \EOS [0]{\spacefactor3000\relax}%
\providecommand \BibitemShut  [1]{\csname bibitem#1\endcsname}%
\let\auto@bib@innerbib\@empty
\bibitem [{\citenamefont {Dutra}\ \emph {et~al.}(2014)\citenamefont {Dutra},
  \citenamefont {Lourenco}, \citenamefont {Avancini}, \citenamefont {Carlson},
  \citenamefont {Delfino}, \citenamefont {Menezes}, \citenamefont
  {Providencia}, \citenamefont {Typel},\ and\ \citenamefont {Stone}}]{RMF-nm}%
  \BibitemOpen
  \bibfield  {author} {\bibinfo {author} {\bibfnamefont {M.}~\bibnamefont
  {Dutra}}, \bibinfo {author} {\bibfnamefont {O.}~\bibnamefont {Lourenco}},
  \bibinfo {author} {\bibfnamefont {S.~S.}\ \bibnamefont {Avancini}}, \bibinfo
  {author} {\bibfnamefont {B.~V.}\ \bibnamefont {Carlson}}, \bibinfo {author}
  {\bibfnamefont {A.}~\bibnamefont {Delfino}}, \bibinfo {author} {\bibfnamefont
  {D.~P.}\ \bibnamefont {Menezes}}, \bibinfo {author} {\bibfnamefont
  {C.}~\bibnamefont {Providencia}}, \bibinfo {author} {\bibfnamefont
  {S.}~\bibnamefont {Typel}}, \ and\ \bibinfo {author} {\bibfnamefont {J.~R.}\
  \bibnamefont {Stone}},\ }\href@noop {} {\bibfield  {journal} {\bibinfo
  {journal} {Phys.\ Rev. C}\ }\textbf {\bibinfo {volume} {90}},\ \bibinfo
  {pages} {055203} (\bibinfo {year} {2014})}\BibitemShut {NoStop}%
\bibitem [{\citenamefont {Nazarewicz}\ \emph {et~al.}(2014)\citenamefont
  {Nazarewicz}, \citenamefont {Reinhard}, \citenamefont {Satula},\ and\
  \citenamefont {Vretenar}}]{NRSV.14}%
  \BibitemOpen
  \bibfield  {author} {\bibinfo {author} {\bibfnamefont {W.}~\bibnamefont
  {Nazarewicz}}, \bibinfo {author} {\bibfnamefont {P.-G.}\ \bibnamefont
  {Reinhard}}, \bibinfo {author} {\bibfnamefont {W.}~\bibnamefont {Satula}}, \
  and\ \bibinfo {author} {\bibfnamefont {D.}~\bibnamefont {Vretenar}},\
  }\href@noop {} {\bibfield  {journal} {\bibinfo  {journal} {Eur.~ Phys.~ J.~
  A}\ }\textbf {\bibinfo {volume} {50}},\ \bibinfo {pages} {20} (\bibinfo
  {year} {2014})}\BibitemShut {NoStop}%
\bibitem [{\citenamefont {Pearson}\ \emph {et~al.}(2014)\citenamefont
  {Pearson}, \citenamefont {Chamel}, \citenamefont {Fantina},\ and\
  \citenamefont {Goriely}}]{PCFG.14}%
  \BibitemOpen
  \bibfield  {author} {\bibinfo {author} {\bibfnamefont {J.~M.}\ \bibnamefont
  {Pearson}}, \bibinfo {author} {\bibfnamefont {N.}~\bibnamefont {Chamel}},
  \bibinfo {author} {\bibfnamefont {A.~F.}\ \bibnamefont {Fantina}}, \ and\
  \bibinfo {author} {\bibfnamefont {S.}~\bibnamefont {Goriely}},\ }\href@noop
  {} {\bibfield  {journal} {\bibinfo  {journal} {Eur.\ Phys. J.}\ }\textbf
  {\bibinfo {volume} {A50}},\ \bibinfo {pages} {43} (\bibinfo {year}
  {2014})}\BibitemShut {NoStop}%
\bibitem [{\citenamefont {Piekarewicz}(2014)}]{P.14}%
  \BibitemOpen
  \bibfield  {author} {\bibinfo {author} {\bibfnamefont {J.}~\bibnamefont
  {Piekarewicz}},\ }\href@noop {} {\bibfield  {journal} {\bibinfo  {journal}
  {Eur.\ Phys. J.}\ }\textbf {\bibinfo {volume} {A50}},\ \bibinfo {pages} {25}
  (\bibinfo {year} {2014})}\BibitemShut {NoStop}%
\bibitem [{\citenamefont {Colo}\ \emph {et~al.}(2014)\citenamefont {Colo},
  \citenamefont {Gard},\ and\ \citenamefont {Sagawa}}]{CGS.14}%
  \BibitemOpen
  \bibfield  {author} {\bibinfo {author} {\bibfnamefont {G.}~\bibnamefont
  {Colo}}, \bibinfo {author} {\bibfnamefont {U.}~\bibnamefont {Gard}}, \ and\
  \bibinfo {author} {\bibfnamefont {H.}~\bibnamefont {Sagawa}},\ }\href@noop {}
  {\bibfield  {journal} {\bibinfo  {journal} {Eur.\ Phys. J.}\ }\textbf
  {\bibinfo {volume} {A50}},\ \bibinfo {pages} {20} (\bibinfo {year}
  {2014})}\BibitemShut {NoStop}%
\bibitem [{\citenamefont {Horowitz}\ \emph {et~al.}(2014)\citenamefont
  {Horowitz}, \citenamefont {Kumar},\ and\ \citenamefont {Michaels}}]{HKM.14}%
  \BibitemOpen
  \bibfield  {author} {\bibinfo {author} {\bibfnamefont {C.~J.}\ \bibnamefont
  {Horowitz}}, \bibinfo {author} {\bibfnamefont {K.~S.}\ \bibnamefont {Kumar}},
  \ and\ \bibinfo {author} {\bibfnamefont {R.}~\bibnamefont {Michaels}},\
  }\href@noop {} {\bibfield  {journal} {\bibinfo  {journal} {Eur.\ Phys. J.}\
  }\textbf {\bibinfo {volume} {A50}},\ \bibinfo {pages} {48} (\bibinfo {year}
  {2014})}\BibitemShut {NoStop}%
\bibitem [{\citenamefont {Bender}\ \emph {et~al.}(2003)\citenamefont {Bender},
  \citenamefont {Heenen},\ and\ \citenamefont {Reinhard}}]{BHP.03}%
  \BibitemOpen
  \bibfield  {author} {\bibinfo {author} {\bibfnamefont {M.}~\bibnamefont
  {Bender}}, \bibinfo {author} {\bibfnamefont {P.-H.}\ \bibnamefont {Heenen}},
  \ and\ \bibinfo {author} {\bibfnamefont {P.-G.}\ \bibnamefont {Reinhard}},\
  }\href@noop {} {\bibfield  {journal} {\bibinfo  {journal} {Rev.\ Mod.\
  Phys.}\ }\textbf {\bibinfo {volume} {75}},\ \bibinfo {pages} {121} (\bibinfo
  {year} {2003})}\BibitemShut {NoStop}%
\bibitem [{\citenamefont {Vretenar}\ \emph {et~al.}(2005)\citenamefont
  {Vretenar}, \citenamefont {Afanasjev}, \citenamefont {Lalazissis},\ and\
  \citenamefont {Ring}}]{VALR.05}%
  \BibitemOpen
  \bibfield  {author} {\bibinfo {author} {\bibfnamefont {D.}~\bibnamefont
  {Vretenar}}, \bibinfo {author} {\bibfnamefont {A.~V.}\ \bibnamefont
  {Afanasjev}}, \bibinfo {author} {\bibfnamefont {G.~A.}\ \bibnamefont
  {Lalazissis}}, \ and\ \bibinfo {author} {\bibfnamefont {P.}~\bibnamefont
  {Ring}},\ }\href@noop {} {\bibfield  {journal} {\bibinfo  {journal} {Phys.\
  Rep.}\ }\textbf {\bibinfo {volume} {409}},\ \bibinfo {pages} {101} (\bibinfo
  {year} {2005})}\BibitemShut {NoStop}%
\bibitem [{\citenamefont {Reinhard}\ and\ \citenamefont
  {Agrawal}(2011)}]{RA.11}%
  \BibitemOpen
  \bibfield  {author} {\bibinfo {author} {\bibfnamefont {P.-G.}\ \bibnamefont
  {Reinhard}}\ and\ \bibinfo {author} {\bibfnamefont {B.~K.}\ \bibnamefont
  {Agrawal}},\ }\href@noop {} {\bibfield  {journal} {\bibinfo  {journal} {Int.\
  Jour.\ Mod.\ Phys.}\ }\textbf {\bibinfo {volume} {E20}},\ \bibinfo {pages}
  {1379} (\bibinfo {year} {2011})}\BibitemShut {NoStop}%
\bibitem [{\citenamefont {Agbemava}\ \emph {et~al.}(2014)\citenamefont
  {Agbemava}, \citenamefont {Afanasjev}, \citenamefont {Ray},\ and\
  \citenamefont {Ring}}]{AARR.14}%
  \BibitemOpen
  \bibfield  {author} {\bibinfo {author} {\bibfnamefont {S.~E.}\ \bibnamefont
  {Agbemava}}, \bibinfo {author} {\bibfnamefont {A.~V.}\ \bibnamefont
  {Afanasjev}}, \bibinfo {author} {\bibfnamefont {D.}~\bibnamefont {Ray}}, \
  and\ \bibinfo {author} {\bibfnamefont {P.}~\bibnamefont {Ring}},\ }\href@noop
  {} {\bibfield  {journal} {\bibinfo  {journal} {Phys.\ Rev. C}\ }\textbf
  {\bibinfo {volume} {89}},\ \bibinfo {pages} {054320} (\bibinfo {year}
  {2014})}\BibitemShut {NoStop}%
\bibitem [{\citenamefont {Dutra}\ \emph {et~al.}(2016)\citenamefont {Dutra},
  \citenamefont {Lourenco},\ and\ \citenamefont {Menezes}}]{DLM.16}%
  \BibitemOpen
  \bibfield  {author} {\bibinfo {author} {\bibfnamefont {M.}~\bibnamefont
  {Dutra}}, \bibinfo {author} {\bibfnamefont {O.}~\bibnamefont {Lourenco}}, \
  and\ \bibinfo {author} {\bibfnamefont {D.~P.}\ \bibnamefont {Menezes}},\
  }\href@noop {} {\bibfield  {journal} {\bibinfo  {journal} {Phys. Rev. C}\
  }\textbf {\bibinfo {volume} {93}},\ \bibinfo {pages} {025806} (\bibinfo
  {year} {2016})}\BibitemShut {NoStop}%
\bibitem [{\citenamefont {Demorest}\ \emph {et~al.}(2010)\citenamefont
  {Demorest}, \citenamefont {Pennucci}, \citenamefont {Ransom}, \citenamefont
  {Roberts},\ and\ \citenamefont {Hessels}}]{NS_mass.1}%
  \BibitemOpen
  \bibfield  {author} {\bibinfo {author} {\bibfnamefont {P.~B.}\ \bibnamefont
  {Demorest}}, \bibinfo {author} {\bibfnamefont {T.}~\bibnamefont {Pennucci}},
  \bibinfo {author} {\bibfnamefont {S.~M.}\ \bibnamefont {Ransom}}, \bibinfo
  {author} {\bibfnamefont {M.~S.~E.}\ \bibnamefont {Roberts}}, \ and\ \bibinfo
  {author} {\bibfnamefont {J.~W.~T.}\ \bibnamefont {Hessels}},\ }\href@noop {}
  {\bibfield  {journal} {\bibinfo  {journal} {Nature}\ }\textbf {\bibinfo
  {volume} {467}},\ \bibinfo {pages} {1081} (\bibinfo {year}
  {2010})}\BibitemShut {NoStop}%
\bibitem [{\citenamefont {Antoniadis}\ \emph {et~al.}(2013)\citenamefont
  {Antoniadis}, \citenamefont {Freire}, \citenamefont {Wex}, \citenamefont
  {Tauris}, \citenamefont {Lynch}, \citenamefont {van Kerkwijk}, \citenamefont
  {Kramer}, \citenamefont {Bassa}, \citenamefont {Dhillon}, \citenamefont
  {Driebe}, \citenamefont {Hessels}, \citenamefont {Kaspi}, \citenamefont
  {Kondratiev}, \citenamefont {Langer}, \citenamefont {Marsh}, \citenamefont
  {McLaughlin}, \citenamefont {Pennucci}, \citenamefont {Ransom}, \citenamefont
  {Stairs}, \citenamefont {van Leeuwen}, \citenamefont {Verbiest},\ and\
  \citenamefont {Whelan}}]{NS_mass.2}%
  \BibitemOpen
  \bibfield  {author} {\bibinfo {author} {\bibfnamefont {J.}~\bibnamefont
  {Antoniadis}}, \bibinfo {author} {\bibfnamefont {P.~C.~C.}\ \bibnamefont
  {Freire}}, \bibinfo {author} {\bibfnamefont {N.}~\bibnamefont {Wex}},
  \bibinfo {author} {\bibfnamefont {T.~M.}\ \bibnamefont {Tauris}}, \bibinfo
  {author} {\bibfnamefont {R.~S.}\ \bibnamefont {Lynch}}, \bibinfo {author}
  {\bibfnamefont {M.~H.}\ \bibnamefont {van Kerkwijk}}, \bibinfo {author}
  {\bibfnamefont {M.}~\bibnamefont {Kramer}}, \bibinfo {author} {\bibfnamefont
  {C.}~\bibnamefont {Bassa}}, \bibinfo {author} {\bibfnamefont {V.~S.}\
  \bibnamefont {Dhillon}}, \bibinfo {author} {\bibfnamefont {T.}~\bibnamefont
  {Driebe}}, \bibinfo {author} {\bibfnamefont {J.~W.~T.}\ \bibnamefont
  {Hessels}}, \bibinfo {author} {\bibfnamefont {V.~M.}\ \bibnamefont {Kaspi}},
  \bibinfo {author} {\bibfnamefont {V.~I.}\ \bibnamefont {Kondratiev}},
  \bibinfo {author} {\bibfnamefont {N.}~\bibnamefont {Langer}}, \bibinfo
  {author} {\bibfnamefont {T.~R.}\ \bibnamefont {Marsh}}, \bibinfo {author}
  {\bibfnamefont {M.~A.}\ \bibnamefont {McLaughlin}}, \bibinfo {author}
  {\bibfnamefont {T.~T.}\ \bibnamefont {Pennucci}}, \bibinfo {author}
  {\bibfnamefont {S.~M.}\ \bibnamefont {Ransom}}, \bibinfo {author}
  {\bibfnamefont {I.~H.}\ \bibnamefont {Stairs}}, \bibinfo {author}
  {\bibfnamefont {J.}~\bibnamefont {van Leeuwen}}, \bibinfo {author}
  {\bibfnamefont {J.~P.~W.}\ \bibnamefont {Verbiest}}, \ and\ \bibinfo {author}
  {\bibfnamefont {D.~G.}\ \bibnamefont {Whelan}},\ }\href@noop {} {\ \textbf
  {\bibinfo {volume} {340}},\ \bibinfo {pages} {448} (\bibinfo {year}
  {2013})}\BibitemShut {NoStop}%
\bibitem [{\citenamefont {Chatterjee}\ and\ \citenamefont
  {Vida{\~n}a}(2016)}]{CV.16}%
  \BibitemOpen
  \bibfield  {author} {\bibinfo {author} {\bibfnamefont {D.}~\bibnamefont
  {Chatterjee}}\ and\ \bibinfo {author} {\bibfnamefont {I.}~\bibnamefont
  {Vida{\~n}a}},\ }\href@noop {} {\bibfield  {journal} {\bibinfo  {journal}
  {Eur.\ Phys. J.}\ }\textbf {\bibinfo {volume} {A52}},\ \bibinfo {pages} {29}
  (\bibinfo {year} {2016})}\BibitemShut {NoStop}%
\bibitem [{\citenamefont {Erler}\ \emph {et~al.}(2012)\citenamefont {Erler},
  \citenamefont {Birge}, \citenamefont {Kortelainen}, \citenamefont
  {Nazarewicz}, \citenamefont {Olsen}, \citenamefont {Perhac},\ and\
  \citenamefont {Stoitsov}}]{Eet.12}%
  \BibitemOpen
  \bibfield  {author} {\bibinfo {author} {\bibfnamefont {J.}~\bibnamefont
  {Erler}}, \bibinfo {author} {\bibfnamefont {N.}~\bibnamefont {Birge}},
  \bibinfo {author} {\bibfnamefont {M.}~\bibnamefont {Kortelainen}}, \bibinfo
  {author} {\bibfnamefont {W.}~\bibnamefont {Nazarewicz}}, \bibinfo {author}
  {\bibfnamefont {E.}~\bibnamefont {Olsen}}, \bibinfo {author} {\bibfnamefont
  {A.~M.}\ \bibnamefont {Perhac}}, \ and\ \bibinfo {author} {\bibfnamefont
  {M.}~\bibnamefont {Stoitsov}},\ }\href@noop {} {\bibfield  {journal}
  {\bibinfo  {journal} {Nature}\ }\textbf {\bibinfo {volume} {486}},\ \bibinfo
  {pages} {509} (\bibinfo {year} {2012})}\BibitemShut {NoStop}%
\bibitem [{\citenamefont {Afanasjev}\ \emph {et~al.}(2013)\citenamefont
  {Afanasjev}, \citenamefont {Agbemava}, \citenamefont {Ray},\ and\
  \citenamefont {Ring}}]{AARR.13}%
  \BibitemOpen
  \bibfield  {author} {\bibinfo {author} {\bibfnamefont {A.~V.}\ \bibnamefont
  {Afanasjev}}, \bibinfo {author} {\bibfnamefont {S.~E.}\ \bibnamefont
  {Agbemava}}, \bibinfo {author} {\bibfnamefont {D.}~\bibnamefont {Ray}}, \
  and\ \bibinfo {author} {\bibfnamefont {P.}~\bibnamefont {Ring}},\ }\href@noop
  {} {\bibfield  {journal} {\bibinfo  {journal} {Phys.\ Lett. B}\ }\textbf
  {\bibinfo {volume} {726}},\ \bibinfo {pages} {680} (\bibinfo {year}
  {2013})}\BibitemShut {NoStop}%
\bibitem [{\citenamefont {Mumpower}\ \emph {et~al.}(2016)\citenamefont
  {Mumpower}, \citenamefont {Surman}, \citenamefont {McLaughlin},\ and\
  \citenamefont {Aprahamian}}]{MSMA.16}%
  \BibitemOpen
  \bibfield  {author} {\bibinfo {author} {\bibfnamefont {M.~R.}\ \bibnamefont
  {Mumpower}}, \bibinfo {author} {\bibfnamefont {R.}~\bibnamefont {Surman}},
  \bibinfo {author} {\bibfnamefont {G.~C.}\ \bibnamefont {McLaughlin}}, \ and\
  \bibinfo {author} {\bibfnamefont {A.}~\bibnamefont {Aprahamian}},\
  }\href@noop {} {\bibfield  {journal} {\bibinfo  {journal} {Prog. Part. Nucl.
  Phys.}\ }\textbf {\bibinfo {volume} {86}},\ \bibinfo {pages} {86 } (\bibinfo
  {year} {2016})}\BibitemShut {NoStop}%
\bibitem [{\citenamefont {Lalazissis}\ \emph {et~al.}(2009)\citenamefont
  {Lalazissis}, \citenamefont {Karatzikos}, \citenamefont {Fossion},
  \citenamefont {Arteaga}, \citenamefont {Afanasjev},\ and\ \citenamefont
  {Ring}}]{NL3*}%
  \BibitemOpen
  \bibfield  {author} {\bibinfo {author} {\bibfnamefont {G.~A.}\ \bibnamefont
  {Lalazissis}}, \bibinfo {author} {\bibfnamefont {S.}~\bibnamefont
  {Karatzikos}}, \bibinfo {author} {\bibfnamefont {R.}~\bibnamefont {Fossion}},
  \bibinfo {author} {\bibfnamefont {D.~P.}\ \bibnamefont {Arteaga}}, \bibinfo
  {author} {\bibfnamefont {A.~V.}\ \bibnamefont {Afanasjev}}, \ and\ \bibinfo
  {author} {\bibfnamefont {P.}~\bibnamefont {Ring}},\ }\href@noop {} {\bibfield
   {journal} {\bibinfo  {journal} {Phys.\ Lett.}\ }\textbf {\bibinfo {volume}
  {B671}},\ \bibinfo {pages} {36} (\bibinfo {year} {2009})}\BibitemShut
  {NoStop}%
\bibitem [{\citenamefont {Lalazissis}\ \emph {et~al.}(2005)\citenamefont
  {Lalazissis}, \citenamefont {Nik{\v{s}}i{\'{c}}}, \citenamefont {Vretenar},\
  and\ \citenamefont {Ring}}]{DD-ME2}%
  \BibitemOpen
  \bibfield  {author} {\bibinfo {author} {\bibfnamefont {G.~A.}\ \bibnamefont
  {Lalazissis}}, \bibinfo {author} {\bibfnamefont {T.}~\bibnamefont
  {Nik{\v{s}}i{\'{c}}}}, \bibinfo {author} {\bibfnamefont {D.}~\bibnamefont
  {Vretenar}}, \ and\ \bibinfo {author} {\bibfnamefont {P.}~\bibnamefont
  {Ring}},\ }\href@noop {} {\bibfield  {journal} {\bibinfo  {journal} {Phys.\
  Rev. C}\ }\textbf {\bibinfo {volume} {71}},\ \bibinfo {pages} {024312}
  (\bibinfo {year} {2005})}\BibitemShut {NoStop}%
\bibitem [{\citenamefont {Nik\v{s}i\'{c}}\ \emph {et~al.}(2008)\citenamefont
  {Nik\v{s}i\'{c}}, \citenamefont {Vretenar},\ and\ \citenamefont
  {Ring}}]{DD-PC1}%
  \BibitemOpen
  \bibfield  {author} {\bibinfo {author} {\bibfnamefont {T.}~\bibnamefont
  {Nik\v{s}i\'{c}}}, \bibinfo {author} {\bibfnamefont {D.}~\bibnamefont
  {Vretenar}}, \ and\ \bibinfo {author} {\bibfnamefont {P.}~\bibnamefont
  {Ring}},\ }\href@noop {} {\bibfield  {journal} {\bibinfo  {journal} {Phys.\
  Rev. C}\ }\textbf {\bibinfo {volume} {78}},\ \bibinfo {pages} {034318}
  (\bibinfo {year} {2008})}\BibitemShut {NoStop}%
\bibitem [{\citenamefont {Roca-Maza}\ \emph {et~al.}(2011)\citenamefont
  {Roca-Maza}, \citenamefont {Vi{\~n}as}, \citenamefont {Centelles},
  \citenamefont {Ring},\ and\ \citenamefont {Schuck}}]{DD-MEdelta}%
  \BibitemOpen
  \bibfield  {author} {\bibinfo {author} {\bibfnamefont {X.}~\bibnamefont
  {Roca-Maza}}, \bibinfo {author} {\bibfnamefont {X.}~\bibnamefont
  {Vi{\~n}as}}, \bibinfo {author} {\bibfnamefont {M.}~\bibnamefont
  {Centelles}}, \bibinfo {author} {\bibfnamefont {P.}~\bibnamefont {Ring}}, \
  and\ \bibinfo {author} {\bibfnamefont {P.}~\bibnamefont {Schuck}},\
  }\href@noop {} {\bibfield  {journal} {\bibinfo  {journal} {Phys.\ Rev. C}\
  }\textbf {\bibinfo {volume} {84}},\ \bibinfo {pages} {054309} (\bibinfo
  {year} {2011})}\BibitemShut {NoStop}%
\bibitem [{\citenamefont {Afanasjev}\ \emph {et~al.}(2015)\citenamefont
  {Afanasjev}, \citenamefont {Agbemava}, \citenamefont {Ray},\ and\
  \citenamefont {Ring}}]{AARR.15}%
  \BibitemOpen
  \bibfield  {author} {\bibinfo {author} {\bibfnamefont {A.~V.}\ \bibnamefont
  {Afanasjev}}, \bibinfo {author} {\bibfnamefont {S.~E.}\ \bibnamefont
  {Agbemava}}, \bibinfo {author} {\bibfnamefont {D.}~\bibnamefont {Ray}}, \
  and\ \bibinfo {author} {\bibfnamefont {P.}~\bibnamefont {Ring}},\ }\href@noop
  {} {\bibfield  {journal} {\bibinfo  {journal} {Phys.\ Rev.\ C}\ }\textbf
  {\bibinfo {volume} {91}},\ \bibinfo {pages} {014324} (\bibinfo {year}
  {2015})}\BibitemShut {NoStop}%
\bibitem [{\citenamefont {Zhao}\ \emph {et~al.}(2010)\citenamefont {Zhao},
  \citenamefont {Li}, \citenamefont {Yao},\ and\ \citenamefont
  {Meng}}]{PC-PK1}%
  \BibitemOpen
  \bibfield  {author} {\bibinfo {author} {\bibfnamefont {P.~W.}\ \bibnamefont
  {Zhao}}, \bibinfo {author} {\bibfnamefont {Z.~P.}\ \bibnamefont {Li}},
  \bibinfo {author} {\bibfnamefont {J.~M.}\ \bibnamefont {Yao}}, \ and\
  \bibinfo {author} {\bibfnamefont {J.}~\bibnamefont {Meng}},\ }\href@noop {}
  {\bibfield  {journal} {\bibinfo  {journal} {Phys.\ Rev. C}\ }\textbf
  {\bibinfo {volume} {82}},\ \bibinfo {pages} {054319} (\bibinfo {year}
  {2010})}\BibitemShut {NoStop}%
\bibitem [{\citenamefont {Zhang}\ \emph {et~al.}(2014)\citenamefont {Zhang},
  \citenamefont {Niu}, \citenamefont {Li}, \citenamefont {Yao},\ and\
  \citenamefont {Meng}}]{ZNLYM.14}%
  \BibitemOpen
  \bibfield  {author} {\bibinfo {author} {\bibfnamefont {Q.~S.}\ \bibnamefont
  {Zhang}}, \bibinfo {author} {\bibfnamefont {Z.~M.}\ \bibnamefont {Niu}},
  \bibinfo {author} {\bibfnamefont {Z.~P.}\ \bibnamefont {Li}}, \bibinfo
  {author} {\bibfnamefont {J.~M.}\ \bibnamefont {Yao}}, \ and\ \bibinfo
  {author} {\bibfnamefont {J.}~\bibnamefont {Meng}},\ }\href@noop {} {\bibfield
   {journal} {\bibinfo  {journal} {Frontiers of Physics}\ }\textbf {\bibinfo
  {volume} {9}},\ \bibinfo {pages} {529} (\bibinfo {year} {2014})}\BibitemShut
  {NoStop}%
\bibitem [{\citenamefont {Lu}\ \emph {et~al.}(2015)\citenamefont {Lu},
  \citenamefont {Li}, \citenamefont {Li}, \citenamefont {Yao},\ and\
  \citenamefont {Meng}}]{LLLYM.15}%
  \BibitemOpen
  \bibfield  {author} {\bibinfo {author} {\bibfnamefont {K.~Q.}\ \bibnamefont
  {Lu}}, \bibinfo {author} {\bibfnamefont {Z.~X.}\ \bibnamefont {Li}}, \bibinfo
  {author} {\bibfnamefont {Z.~P.}\ \bibnamefont {Li}}, \bibinfo {author}
  {\bibfnamefont {J.~M.}\ \bibnamefont {Yao}}, \ and\ \bibinfo {author}
  {\bibfnamefont {J.}~\bibnamefont {Meng}},\ }\href {\doibase
  10.1103/PhysRevC.91.027304} {\bibfield  {journal} {\bibinfo  {journal} {Phys.
  Rev. C}\ }\textbf {\bibinfo {volume} {91}},\ \bibinfo {pages} {027304}
  (\bibinfo {year} {2015})}\BibitemShut {NoStop}%
\bibitem [{\citenamefont {Akmal}\ \emph {et~al.}(1998)\citenamefont {Akmal},
  \citenamefont {Pandharipande},\ and\ \citenamefont {Ravenhall}}]{APR.98}%
  \BibitemOpen
  \bibfield  {author} {\bibinfo {author} {\bibfnamefont {A.}~\bibnamefont
  {Akmal}}, \bibinfo {author} {\bibfnamefont {V.~R.}\ \bibnamefont
  {Pandharipande}}, \ and\ \bibinfo {author} {\bibfnamefont {D.~G.}\
  \bibnamefont {Ravenhall}},\ }\href {\doibase 10.1103/PhysRevC.58.1804}
  {\bibfield  {journal} {\bibinfo  {journal} {Phys. Rev. C}\ }\textbf {\bibinfo
  {volume} {58}},\ \bibinfo {pages} {1804} (\bibinfo {year}
  {1998})}\BibitemShut {NoStop}%
\bibitem [{\citenamefont {Li}\ and\ \citenamefont {Schulze}(2008)}]{LS.08}%
  \BibitemOpen
  \bibfield  {author} {\bibinfo {author} {\bibfnamefont {Z.~H.}\ \bibnamefont
  {Li}}\ and\ \bibinfo {author} {\bibfnamefont {H.-J.}\ \bibnamefont
  {Schulze}},\ }\href {\doibase 10.1103/PhysRevC.78.028801} {\bibfield
  {journal} {\bibinfo  {journal} {Phys. Rev. C}\ }\textbf {\bibinfo {volume}
  {78}},\ \bibinfo {pages} {028801} (\bibinfo {year} {2008})}\BibitemShut
  {NoStop}%
\bibitem [{\citenamefont {Wang}\ \emph {et~al.}(2012)\citenamefont {Wang},
  \citenamefont {Audi}, \citenamefont {Wapstra}, \citenamefont {Kondev},
  \citenamefont {MacCormick}, \citenamefont {Xu},\ and\ \citenamefont
  {Pfeiffer}}]{AME2012}%
  \BibitemOpen
  \bibfield  {author} {\bibinfo {author} {\bibfnamefont {M.}~\bibnamefont
  {Wang}}, \bibinfo {author} {\bibfnamefont {G.}~\bibnamefont {Audi}}, \bibinfo
  {author} {\bibfnamefont {A.~H.}\ \bibnamefont {Wapstra}}, \bibinfo {author}
  {\bibfnamefont {F.~G.}\ \bibnamefont {Kondev}}, \bibinfo {author}
  {\bibfnamefont {M.}~\bibnamefont {MacCormick}}, \bibinfo {author}
  {\bibfnamefont {X.}~\bibnamefont {Xu}}, \ and\ \bibinfo {author}
  {\bibfnamefont {B.}~\bibnamefont {Pfeiffer}},\ }\href@noop {} {\bibfield
  {journal} {\bibinfo  {journal} {Chinese Physics}\ }\textbf {\bibinfo {volume}
  {C36}},\ \bibinfo {pages} {1603} (\bibinfo {year} {2012})}\BibitemShut
  {NoStop}%
\bibitem [{\citenamefont {Schatz}(2014)}]{S-priv.14}%
  \BibitemOpen
  \bibfield  {author} {\bibinfo {author} {\bibfnamefont {H.}~\bibnamefont
  {Schatz}},\ }\href@noop {} {\bibfield  {journal} {\bibinfo  {journal}
  {private communication, see also
  https://groups.nscl.msu.edu/frib/rates/fribrates.html}\ } (\bibinfo {year}
  {2014})}\BibitemShut {NoStop}%
\bibitem [{\citenamefont {McDonnell}\ \emph {et~al.}(2015)\citenamefont
  {McDonnell}, \citenamefont {Schunck}, \citenamefont {Higdon}, \citenamefont
  {Sarich}, \citenamefont {Wild},\ and\ \citenamefont
  {Nazarewicz}}]{MSHSWN.15}%
  \BibitemOpen
  \bibfield  {author} {\bibinfo {author} {\bibfnamefont {J.~D.}\ \bibnamefont
  {McDonnell}}, \bibinfo {author} {\bibfnamefont {N.}~\bibnamefont {Schunck}},
  \bibinfo {author} {\bibfnamefont {D.}~\bibnamefont {Higdon}}, \bibinfo
  {author} {\bibfnamefont {J.}~\bibnamefont {Sarich}}, \bibinfo {author}
  {\bibfnamefont {S.~M.}\ \bibnamefont {Wild}}, \ and\ \bibinfo {author}
  {\bibfnamefont {W.}~\bibnamefont {Nazarewicz}},\ }\href {\doibase
  10.1103/PhysRevLett.114.122501} {\bibfield  {journal} {\bibinfo  {journal}
  {Phys. Rev. Lett.}\ }\textbf {\bibinfo {volume} {114}},\ \bibinfo {pages}
  {122501} (\bibinfo {year} {2015})}\BibitemShut {NoStop}%
\bibitem [{\citenamefont {Agbemava}\ \emph {et~al.}(2015)\citenamefont
  {Agbemava}, \citenamefont {Afanasjev}, \citenamefont {Nakatsukasa},\ and\
  \citenamefont {Ring}}]{AANR.15}%
  \BibitemOpen
  \bibfield  {author} {\bibinfo {author} {\bibfnamefont {S.~E.}\ \bibnamefont
  {Agbemava}}, \bibinfo {author} {\bibfnamefont {A.~V.}\ \bibnamefont
  {Afanasjev}}, \bibinfo {author} {\bibfnamefont {T.}~\bibnamefont
  {Nakatsukasa}}, \ and\ \bibinfo {author} {\bibfnamefont {P.}~\bibnamefont
  {Ring}},\ }\href {\doibase 10.1103/PhysRevC.92.054310} {\bibfield  {journal}
  {\bibinfo  {journal} {Phys. Rev. C}\ }\textbf {\bibinfo {volume} {92}},\
  \bibinfo {pages} {054310} (\bibinfo {year} {2015})}\BibitemShut {NoStop}%
\bibitem [{\citenamefont {Agbemava}\ \emph {et~al.}(2016)\citenamefont
  {Agbemava}, \citenamefont {Afanasjev},\ and\ \citenamefont {Ring}}]{AAR.16}%
  \BibitemOpen
  \bibfield  {author} {\bibinfo {author} {\bibfnamefont {S.~E.}\ \bibnamefont
  {Agbemava}}, \bibinfo {author} {\bibfnamefont {A.~V.}\ \bibnamefont
  {Afanasjev}}, \ and\ \bibinfo {author} {\bibfnamefont {P.}~\bibnamefont
  {Ring}},\ }\href@noop {} {\bibfield  {journal} {\bibinfo  {journal} {Phys.
  Rev. C}\ }\textbf {\bibinfo {volume} {93}},\ \bibinfo {pages} {044304}
  (\bibinfo {year} {2016})}\BibitemShut {NoStop}%
\bibitem [{\citenamefont {Reinhard}\ and\ \citenamefont
  {Nazarewicz}(2016)}]{RN.16}%
  \BibitemOpen
  \bibfield  {author} {\bibinfo {author} {\bibfnamefont {P.-G.}\ \bibnamefont
  {Reinhard}}\ and\ \bibinfo {author} {\bibfnamefont {W.}~\bibnamefont
  {Nazarewicz}},\ }\href@noop {} {\bibfield  {journal} {\bibinfo  {journal}
  {arXiv: nucl-th/1601.06324v1}\ } (\bibinfo {year} {2016})}\BibitemShut
  {NoStop}%
\bibitem [{\citenamefont {Fattoyev}\ and\ \citenamefont
  {Piekarewicz}(2011)}]{FP.11}%
  \BibitemOpen
  \bibfield  {author} {\bibinfo {author} {\bibfnamefont {F.~J.}\ \bibnamefont
  {Fattoyev}}\ and\ \bibinfo {author} {\bibfnamefont {J.}~\bibnamefont
  {Piekarewicz}},\ }\href@noop {} {\bibfield  {journal} {\bibinfo  {journal}
  {Phys.\ Rev. C}\ }\textbf {\bibinfo {volume} {84}},\ \bibinfo {pages}
  {064302} (\bibinfo {year} {2011})}\BibitemShut {NoStop}%
\bibitem [{\citenamefont {Roca-Maza}\ \emph {et~al.}(2015)\citenamefont
  {Roca-Maza}, \citenamefont {Paar},\ and\ \citenamefont {Col\`o}}]{RNC.15}%
  \BibitemOpen
  \bibfield  {author} {\bibinfo {author} {\bibfnamefont {X.}~\bibnamefont
  {Roca-Maza}}, \bibinfo {author} {\bibfnamefont {N.}~\bibnamefont {Paar}}, \
  and\ \bibinfo {author} {\bibfnamefont {G.}~\bibnamefont {Col\`o}},\ }\href
  {http://stacks.iop.org/0954-3899/42/i=3/a=034033} {\bibfield  {journal}
  {\bibinfo  {journal} {J.~ Phys. G}\ }\textbf {\bibinfo {volume} {42}},\
  \bibinfo {pages} {034033} (\bibinfo {year} {2015})}\BibitemShut {NoStop}%
\bibitem [{\citenamefont {Reinhard}\ and\ \citenamefont
  {Nazarewicz}(2010)}]{RN.10}%
  \BibitemOpen
  \bibfield  {author} {\bibinfo {author} {\bibfnamefont {P.~G.}\ \bibnamefont
  {Reinhard}}\ and\ \bibinfo {author} {\bibfnamefont {W.}~\bibnamefont
  {Nazarewicz}},\ }\href@noop {} {\bibfield  {journal} {\bibinfo  {journal}
  {Phys.\ Rev. C}\ }\textbf {\bibinfo {volume} {81}},\ \bibinfo {pages}
  {051303(R)} (\bibinfo {year} {2010})}\BibitemShut {NoStop}%
\bibitem [{\citenamefont {Piekarewicz}\ \emph {et~al.}(2012)\citenamefont
  {Piekarewicz}, \citenamefont {Agrawal}, \citenamefont {Col\`o}, \citenamefont
  {Nazarewicz}, \citenamefont {Paar}, \citenamefont {Reinhard}, \citenamefont
  {Roca-Maza},\ and\ \citenamefont {Vretenar}}]{PACNPRRV.12}%
  \BibitemOpen
  \bibfield  {author} {\bibinfo {author} {\bibfnamefont {J.}~\bibnamefont
  {Piekarewicz}}, \bibinfo {author} {\bibfnamefont {B.~K.}\ \bibnamefont
  {Agrawal}}, \bibinfo {author} {\bibfnamefont {G.}~\bibnamefont {Col\`o}},
  \bibinfo {author} {\bibfnamefont {W.}~\bibnamefont {Nazarewicz}}, \bibinfo
  {author} {\bibfnamefont {N.}~\bibnamefont {Paar}}, \bibinfo {author}
  {\bibfnamefont {P.-G.}\ \bibnamefont {Reinhard}}, \bibinfo {author}
  {\bibfnamefont {X.}~\bibnamefont {Roca-Maza}}, \ and\ \bibinfo {author}
  {\bibfnamefont {D.}~\bibnamefont {Vretenar}},\ }\href {\doibase
  10.1103/PhysRevC.85.041302} {\bibfield  {journal} {\bibinfo  {journal} {Phys.
  Rev. C}\ }\textbf {\bibinfo {volume} {85}},\ \bibinfo {pages} {041302}
  (\bibinfo {year} {2012})}\BibitemShut {NoStop}%
\bibitem [{\citenamefont {Jaminon}\ and\ \citenamefont {Mahaux}(1989)}]{JM.89}%
  \BibitemOpen
  \bibfield  {author} {\bibinfo {author} {\bibfnamefont {M.}~\bibnamefont
  {Jaminon}}\ and\ \bibinfo {author} {\bibfnamefont {C.}~\bibnamefont
  {Mahaux}},\ }\href@noop {} {\bibfield  {journal} {\bibinfo  {journal} {Phys.
  Rev. C}\ }\textbf {\bibinfo {volume} {40}},\ \bibinfo {pages} {354} (\bibinfo
  {year} {1989})}\BibitemShut {NoStop}%
\bibitem [{\citenamefont {Todd-Rutel}\ and\ \citenamefont
  {Piekarewicz}(2005)}]{FSUGold}%
  \BibitemOpen
  \bibfield  {author} {\bibinfo {author} {\bibfnamefont {B.~G.}\ \bibnamefont
  {Todd-Rutel}}\ and\ \bibinfo {author} {\bibfnamefont {J.}~\bibnamefont
  {Piekarewicz}},\ }\href@noop {} {\bibfield  {journal} {\bibinfo  {journal}
  {Phys.\ Rev.\ Lett.}\ }\textbf {\bibinfo {volume} {95}},\ \bibinfo {pages}
  {122501} (\bibinfo {year} {2005})}\BibitemShut {NoStop}%
\bibitem [{\citenamefont {Agbemava}\ \emph {et~al.}(tion)\citenamefont
  {Agbemava}, \citenamefont {Afanasjev}, \citenamefont {Nakatsukasa},\ and\
  \citenamefont {Ring}}]{AANR.16}%
  \BibitemOpen
  \bibfield  {author} {\bibinfo {author} {\bibfnamefont {S.~E.}\ \bibnamefont
  {Agbemava}}, \bibinfo {author} {\bibfnamefont {A.~V.}\ \bibnamefont
  {Afanasjev}}, \bibinfo {author} {\bibfnamefont {T.}~\bibnamefont
  {Nakatsukasa}}, \ and\ \bibinfo {author} {\bibfnamefont {P.}~\bibnamefont
  {Ring}},\ }\href@noop {} {\  (\bibinfo {year} {in preparation})}\BibitemShut
  {NoStop}%
\bibitem [{\citenamefont {Abusara}\ \emph {et~al.}(2010)\citenamefont
  {Abusara}, \citenamefont {Afanasjev},\ and\ \citenamefont {Ring}}]{AAR.10}%
  \BibitemOpen
  \bibfield  {author} {\bibinfo {author} {\bibfnamefont {H.}~\bibnamefont
  {Abusara}}, \bibinfo {author} {\bibfnamefont {A.~V.}\ \bibnamefont
  {Afanasjev}}, \ and\ \bibinfo {author} {\bibfnamefont {P.}~\bibnamefont
  {Ring}},\ }\href@noop {} {\bibfield  {journal} {\bibinfo  {journal} {Phys.\
  Rev. C}\ }\textbf {\bibinfo {volume} {82}},\ \bibinfo {pages} {044303}
  (\bibinfo {year} {2010})}\BibitemShut {NoStop}%
\bibitem [{\citenamefont {Afanasjev}\ and\ \citenamefont
  {Abdurazakov}(2013)}]{AO.13}%
  \BibitemOpen
  \bibfield  {author} {\bibinfo {author} {\bibfnamefont {A.~V.}\ \bibnamefont
  {Afanasjev}}\ and\ \bibinfo {author} {\bibfnamefont {O.}~\bibnamefont
  {Abdurazakov}},\ }\href@noop {} {\bibfield  {journal} {\bibinfo  {journal}
  {Phys.\ Rev. C}\ }\textbf {\bibinfo {volume} {88}},\ \bibinfo {pages}
  {014320} (\bibinfo {year} {2013})}\BibitemShut {NoStop}%
\bibitem [{\citenamefont {Afanasjev}\ and\ \citenamefont
  {Shawaqfeh}(2011)}]{AS.11}%
  \BibitemOpen
  \bibfield  {author} {\bibinfo {author} {\bibfnamefont {A.~V.}\ \bibnamefont
  {Afanasjev}}\ and\ \bibinfo {author} {\bibfnamefont {S.}~\bibnamefont
  {Shawaqfeh}},\ }\href@noop {} {\bibfield  {journal} {\bibinfo  {journal}
  {Phys.\ Lett. B}\ }\textbf {\bibinfo {volume} {706}},\ \bibinfo {pages} {177}
  (\bibinfo {year} {2011})}\BibitemShut {NoStop}%
\bibitem [{\citenamefont {Meng}\ \emph {et~al.}(2013)\citenamefont {Meng},
  \citenamefont {Peng}, \citenamefont {Zhang},\ and\ \citenamefont
  {Zhao}}]{Meng2013Front.Phys.55}%
  \BibitemOpen
  \bibfield  {author} {\bibinfo {author} {\bibfnamefont {J.}~\bibnamefont
  {Meng}}, \bibinfo {author} {\bibfnamefont {J.}~\bibnamefont {Peng}}, \bibinfo
  {author} {\bibfnamefont {S.-Q.}\ \bibnamefont {Zhang}}, \ and\ \bibinfo
  {author} {\bibfnamefont {P.-W.}\ \bibnamefont {Zhao}},\ }\href {\doibase
  10.1007/s11467-013-0287-y} {\bibfield  {journal} {\bibinfo  {journal} {Front.
  Phys.}\ }\textbf {\bibinfo {volume} {8}},\ \bibinfo {pages} {55} (\bibinfo
  {year} {2013})}\BibitemShut {NoStop}%
\bibitem [{\citenamefont {Lu}\ \emph {et~al.}(2014)\citenamefont {Lu},
  \citenamefont {Zhao}, \citenamefont {Zhao},\ and\ \citenamefont
  {Zhou}}]{LZZ.14}%
  \BibitemOpen
  \bibfield  {author} {\bibinfo {author} {\bibfnamefont {B.-N.}\ \bibnamefont
  {Lu}}, \bibinfo {author} {\bibfnamefont {J.}~\bibnamefont {Zhao}}, \bibinfo
  {author} {\bibfnamefont {E.-G.}\ \bibnamefont {Zhao}}, \ and\ \bibinfo
  {author} {\bibfnamefont {S.-G.}\ \bibnamefont {Zhou}},\ }\href {\doibase
  10.1103/PhysRevC.89.014323} {\bibfield  {journal} {\bibinfo  {journal} {Phys.
  Rev. C}\ }\textbf {\bibinfo {volume} {89}},\ \bibinfo {pages} {014323}
  (\bibinfo {year} {2014})}\BibitemShut {NoStop}%
\bibitem [{\citenamefont {Prassa}\ \emph {et~al.}(2012)\citenamefont {Prassa},
  \citenamefont {Nik\ifmmode \check{s}\else \v{s}\fi{}i\ifmmode~\acute{c}\else
  \'{c}\fi{}}, \citenamefont {Lalazissis},\ and\ \citenamefont
  {Vretenar}}]{PNLV.12}%
  \BibitemOpen
  \bibfield  {author} {\bibinfo {author} {\bibfnamefont {V.}~\bibnamefont
  {Prassa}}, \bibinfo {author} {\bibfnamefont {T.}~\bibnamefont {Nik\ifmmode
  \check{s}\else \v{s}\fi{}i\ifmmode~\acute{c}\else \'{c}\fi{}}}, \bibinfo
  {author} {\bibfnamefont {G.~A.}\ \bibnamefont {Lalazissis}}, \ and\ \bibinfo
  {author} {\bibfnamefont {D.}~\bibnamefont {Vretenar}},\ }\href@noop {}
  {\bibfield  {journal} {\bibinfo  {journal} {Phys. Rev. C}\ }\textbf {\bibinfo
  {volume} {86}},\ \bibinfo {pages} {024317} (\bibinfo {year}
  {2012})}\BibitemShut {NoStop}%
\bibitem [{\citenamefont {Stevenson}\ \emph {et~al.}(2013)\citenamefont
  {Stevenson}, \citenamefont {Goddard}, \citenamefont {Stone},\ and\
  \citenamefont {Dutra}}]{SGSD.13}%
  \BibitemOpen
  \bibfield  {author} {\bibinfo {author} {\bibfnamefont {P.~D.}\ \bibnamefont
  {Stevenson}}, \bibinfo {author} {\bibfnamefont {P.~M.}\ \bibnamefont
  {Goddard}}, \bibinfo {author} {\bibfnamefont {J.~R.}\ \bibnamefont {Stone}},
  \ and\ \bibinfo {author} {\bibfnamefont {M.}~\bibnamefont {Dutra}},\
  }\href@noop {} {\bibfield  {journal} {\bibinfo  {journal} {AIP Conf. Proc.}\
  }\textbf {\bibinfo {volume} {1529}},\ \bibinfo {pages} {269} (\bibinfo {year}
  {2013})}\BibitemShut {NoStop}%
\bibitem [{\citenamefont {Angeli}\ and\ \citenamefont
  {Marinova}(2013)}]{AM.13}%
  \BibitemOpen
  \bibfield  {author} {\bibinfo {author} {\bibfnamefont {I.}~\bibnamefont
  {Angeli}}\ and\ \bibinfo {author} {\bibfnamefont {K.~P.}\ \bibnamefont
  {Marinova}},\ }\href@noop {} {\bibfield  {journal} {\bibinfo  {journal} {At.\
  Data Nucl.\ Data Tables}\ }\textbf {\bibinfo {volume} {99}},\ \bibinfo
  {pages} {69} (\bibinfo {year} {2013})}\BibitemShut {NoStop}%
\bibitem [{\citenamefont {Bender}\ \emph {et~al.}(2006)\citenamefont {Bender},
  \citenamefont {Bertsch},\ and\ \citenamefont {Heenen}}]{BBH.06}%
  \BibitemOpen
  \bibfield  {author} {\bibinfo {author} {\bibfnamefont {M.}~\bibnamefont
  {Bender}}, \bibinfo {author} {\bibfnamefont {G.~F.}\ \bibnamefont {Bertsch}},
  \ and\ \bibinfo {author} {\bibfnamefont {P.-H.}\ \bibnamefont {Heenen}},\
  }\href {\doibase 10.1103/PhysRevC.73.034322} {\bibfield  {journal} {\bibinfo
  {journal} {Phys. Rev. C}\ }\textbf {\bibinfo {volume} {73}},\ \bibinfo
  {pages} {034322} (\bibinfo {year} {2006})}\BibitemShut {NoStop}%
\bibitem [{\citenamefont {Goriely}\ \emph {et~al.}(2009)\citenamefont
  {Goriely}, \citenamefont {Chamel},\ and\ \citenamefont {Pearson}}]{GCP.09}%
  \BibitemOpen
  \bibfield  {author} {\bibinfo {author} {\bibfnamefont {S.}~\bibnamefont
  {Goriely}}, \bibinfo {author} {\bibfnamefont {N.}~\bibnamefont {Chamel}}, \
  and\ \bibinfo {author} {\bibfnamefont {J.~M.}\ \bibnamefont {Pearson}},\
  }\href@noop {} {\bibfield  {journal} {\bibinfo  {journal} {Phys.\ Rev.\
  Lett.}\ }\textbf {\bibinfo {volume} {102}},\ \bibinfo {pages} {152503}
  (\bibinfo {year} {2009})}\BibitemShut {NoStop}%
\bibitem [{\citenamefont {Ring}(1996)}]{Ring1996_PPNP37-193}%
  \BibitemOpen
  \bibfield  {author} {\bibinfo {author} {\bibfnamefont {P.}~\bibnamefont
  {Ring}},\ }\href@noop {} {\bibfield  {journal} {\bibinfo  {journal} {Prog.
  Part. Nucl. Phys.}\ }\textbf {\bibinfo {volume} {37}} (\bibinfo {year}
  {1996})}\BibitemShut {NoStop}%
\bibitem [{\citenamefont {Geng}\ \emph {et~al.}(2005)\citenamefont {Geng},
  \citenamefont {Toki},\ and\ \citenamefont {Meng}}]{GTM.05}%
  \BibitemOpen
  \bibfield  {author} {\bibinfo {author} {\bibfnamefont {L.}~\bibnamefont
  {Geng}}, \bibinfo {author} {\bibfnamefont {H.}~\bibnamefont {Toki}}, \ and\
  \bibinfo {author} {\bibfnamefont {J.}~\bibnamefont {Meng}},\ }\href@noop {}
  {\bibfield  {journal} {\bibinfo  {journal} {Prog.\ Theor.\ Phys.}\ }\textbf
  {\bibinfo {volume} {113}},\ \bibinfo {pages} {785} (\bibinfo {year}
  {2005})}\BibitemShut {NoStop}%
\bibitem [{\citenamefont {Bertozzi}\ \emph {et~al.}(1972)\citenamefont
  {Bertozzi}, \citenamefont {Friar}, \citenamefont {Heisenberg},\ and\
  \citenamefont {Negele}}]{BFHN.72}%
  \BibitemOpen
  \bibfield  {author} {\bibinfo {author} {\bibfnamefont {W.}~\bibnamefont
  {Bertozzi}}, \bibinfo {author} {\bibfnamefont {J.}~\bibnamefont {Friar}},
  \bibinfo {author} {\bibfnamefont {J.}~\bibnamefont {Heisenberg}}, \ and\
  \bibinfo {author} {\bibfnamefont {J.~W.}\ \bibnamefont {Negele}},\
  }\href@noop {} {\bibfield  {journal} {\bibinfo  {journal} {Phys.\ Lett. B}\
  }\textbf {\bibinfo {volume} {41}},\ \bibinfo {pages} {408} (\bibinfo {year}
  {1972})}\BibitemShut {NoStop}%
\bibitem [{\citenamefont {Nishimura}\ and\ \citenamefont
  {Sprung}(1987)}]{NS.87}%
  \BibitemOpen
  \bibfield  {author} {\bibinfo {author} {\bibfnamefont {M.}~\bibnamefont
  {Nishimura}}\ and\ \bibinfo {author} {\bibfnamefont {D.~W.~L.}\ \bibnamefont
  {Sprung}},\ }\href@noop {} {\bibfield  {journal} {\bibinfo  {journal} {Prog.\
  Theor.\ Phys.}\ }\textbf {\bibinfo {volume} {77}},\ \bibinfo {pages} {781}
  (\bibinfo {year} {1987})}\BibitemShut {NoStop}%
\bibitem [{\citenamefont {Abusara}\ \emph {et~al.}(2012)\citenamefont
  {Abusara}, \citenamefont {Afanasjev},\ and\ \citenamefont {Ring}}]{AAR.12}%
  \BibitemOpen
  \bibfield  {author} {\bibinfo {author} {\bibfnamefont {H.}~\bibnamefont
  {Abusara}}, \bibinfo {author} {\bibfnamefont {A.~V.}\ \bibnamefont
  {Afanasjev}}, \ and\ \bibinfo {author} {\bibfnamefont {P.}~\bibnamefont
  {Ring}},\ }\href@noop {} {\bibfield  {journal} {\bibinfo  {journal} {Phys.\
  Rev. C}\ }\textbf {\bibinfo {volume} {85}},\ \bibinfo {pages} {024314}
  (\bibinfo {year} {2012})}\BibitemShut {NoStop}%
\bibitem [{\citenamefont {Litvinova}\ and\ \citenamefont
  {Afanasjev}(2011)}]{LA.11}%
  \BibitemOpen
  \bibfield  {author} {\bibinfo {author} {\bibfnamefont {E.~V.}\ \bibnamefont
  {Litvinova}}\ and\ \bibinfo {author} {\bibfnamefont {A.~V.}\ \bibnamefont
  {Afanasjev}},\ }\href@noop {} {\bibfield  {journal} {\bibinfo  {journal}
  {Phys.\ Rev. C}\ }\textbf {\bibinfo {volume} {84}},\ \bibinfo {pages}
  {014305} (\bibinfo {year} {2011})}\BibitemShut {NoStop}%
\bibitem [{\citenamefont {Afanasjev}\ and\ \citenamefont
  {Litvinova}(2015)}]{AL.15}%
  \BibitemOpen
  \bibfield  {author} {\bibinfo {author} {\bibfnamefont {A.~V.}\ \bibnamefont
  {Afanasjev}}\ and\ \bibinfo {author} {\bibfnamefont {E.}~\bibnamefont
  {Litvinova}},\ }\href {\doibase 10.1103/PhysRevC.92.044317} {\bibfield
  {journal} {\bibinfo  {journal} {Phys. Rev. C}\ }\textbf {\bibinfo {volume}
  {92}},\ \bibinfo {pages} {044317} (\bibinfo {year} {2015})}\BibitemShut
  {NoStop}%
\bibitem [{\citenamefont {Dobaczewski}\ \emph {et~al.}(2015)\citenamefont
  {Dobaczewski}, \citenamefont {Afanasjev}, \citenamefont {Bender},
  \citenamefont {Robledo},\ and\ \citenamefont {Shi}}]{DABRS.15}%
  \BibitemOpen
  \bibfield  {author} {\bibinfo {author} {\bibfnamefont {J.}~\bibnamefont
  {Dobaczewski}}, \bibinfo {author} {\bibfnamefont {A.~V.}\ \bibnamefont
  {Afanasjev}}, \bibinfo {author} {\bibfnamefont {M.}~\bibnamefont {Bender}},
  \bibinfo {author} {\bibfnamefont {L.~M.}\ \bibnamefont {Robledo}}, \ and\
  \bibinfo {author} {\bibfnamefont {Y.}~\bibnamefont {Shi}},\ }\href {\doibase
  http://dx.doi.org/10.1016/j.nuclphysa.2015.07.015} {\bibfield  {journal}
  {\bibinfo  {journal} {Nucl. Phys. A}\ }\textbf {\bibinfo {volume} {944}},\
  \bibinfo {pages} {388 } (\bibinfo {year} {2015})}\BibitemShut {NoStop}%
\end{thebibliography}%

\end{document}